\documentclass[a4paper,usenatbib]{mnras}

\makeatletter

\newcommand{\Rmnum}[1]{\expandafter\@slowromancap\romannumeral #1@}
\newcommand{\lsim}{\lesssim}
\newcommand{\gsim}{\lower0.6ex\vbox{\hbox{$ \buildrel{\textstyle >}\over{\sim}\ $}}}
\makeatother

\usepackage{newtxtext,newtxmath}
\usepackage{graphicx}	
\usepackage{amsmath}	
\usepackage{amssymb}	
\usepackage{subcaption}
\usepackage{color}

\newcommand{\yueying}[1]{\textcolor{black}{#1}}

\title[Hydrodynamic simulations of Fuzzy Dark Matter using MP-Gadget]{
Predictions for the Abundance
of High-redshift Galaxies in a
Fuzzy Dark Matter Universe}

\author[Y. Ni et al.]
{Yueying Ni$^{1}$\thanks{Email:yueyingn@andrew.cmu.edu}, Mei-Yu Wang$^{1}$, Yu Feng$^{2}$, Tiziana Di Matteo$^{1,3}$ \\
$^1$McWilliams Center for Cosmology, Department of Physics, Carnegie Mellon University, Pittsburgh, PA 15213 \\
$^2$Berkeley Center for Cosmological Physics, University of California, Berkeley, Berkeley CA, 94720 \\
$^3$ School of Physics,
The University of Melbourne
VIC 3010 Australia}

\date{Accepted XXX. Received YYY; in original form ZZZ}

\pubyear{2019}

\begin{document}
\label{firstpage}
\pagerange{\pageref{firstpage}--\pageref{lastpage}}
\maketitle

\begin{abstract}
During the last decades, rapid progress has been made in measurements of the rest-frame ultraviolet (UV) luminosity function (LF) for high-redshift galaxies ($z \geq 6$). The faint-end of the galaxy LF at these redshifts
provides powerful constraints on different dark matter models that suppress small-scale structure formation. 
In this work we perform full hydrodynamical cosmological simulations of galaxy formation using an alternative DM model composed of extremely light bosonic particles ($m \sim 10^{-22}$ eV), also known as fuzzy dark matter (FDM), and examine the predictions for the galaxy stellar mass function and luminosity function at $z \geq 6$ for a range of FDM masses.
We find that for FDM models with bosonic mass $m = 5\times10^{-22}$ eV, the number density of galaxies with stellar mass $\rm M_* \sim 10^7 M_{\odot}$ is suppressed by $\sim 40\%$ at z = 9, $\sim 20\%$ at z = 5, and the UV LFs within magnitude range of -16 < $M_{\rm UV}$ < -14 is suppressed by $\sim 60\%$ at $z = 9$, $\sim 20\%$ at $z = 5$ comparing to the CDM counterpart simulation.
Comparing our predictions with current measurements of the faint-end LFs ($-18 \leqslant M_{\rm UV} \leqslant -14$), we find that
FDM models with $m_{22} < 5\times10^{-22}$ are ruled out at $3\sigma$ confidence level. 
We expect that future LF measurements by James Webb Space Telescope (JWST), which will extend down to $M_{\rm UV} \sim -13$ for $z \lesssim 10$, with a survey volume that is comparable to the Hubble Ultra Deep Field (HUDF) would have the capability to constrain FDM models to $m\; \gtrsim 10^{-21}$ eV. 

\end{abstract}





\section{Introduction}
\label{section1:introduction}

In the standard paradigm,  hierarchical structure formation arises from the gravitational growth and collapse of the initial dark matter (DM) inhomogeneities.
Owing to its success in describing a range of phenomena over the last several decades, the cold dark matter (CDM) has become the standard description for the formation of cosmic structure and galaxies \citep{White_etal78,Blumenthal_etal84}. In particular, the CDM model is consistent with the cosmic microwave background (CMB) anisotropy spectrum measurements \citep[e.g.][]{Planck2016} as well as observations of the large-scale galaxy clustering measured by wide-field imaging surveys \citep[e.g.][]{Alam2017}.

However, despite the success in predicting structure formation at large-scale, the CDM paradigm has long been challenged by problems at small scales. These include the "cusp-core problem" \citep{Moore1994,Flores1994}, the "missing satellite problem" \citep{Moore1999}, and more recently the "Too Big to Fail problem" \citep{Boylan2011}. 

There are two typical approaches to address these problems. One is to 
appeal to astrophysical process, 
and baryon physics, which can alter small-scale structure \citep[see e.g.][]{Bosch2000,Schaller2015a}.
Another is to diversify the DM properties and introduce alternative DM models that suppress small-scale structure growth. These include warm dark matter (WDM) \citep{Bode2001,Abazajian2006}, decaying dark matter (DDM) \citep{Wang2014,Cheng2015}, self-interacting dark matter (SIDM) \citep{Colin2002}, and fuzzy dark matter (FDM) \citep{Hu2000}.

As an intriguing alternative to CDM, FDM models, which are motivated by axions generic in string theory \citep{Witten1984,Svrcek2006,Cicoli2012} or non-QCD axion mechanism \citep{Chiueh2014}, describe the DM particles in the form of ultra-light scalar bosons. 
In this scenario, a DM particle has an extremely light mass of $m \sim 10^{-22}$ eV so that it exhibits wave nature on astrophysical scales with a de Broglie wavelength in the order of $\sim 1~\rm kpc$ \citep{Hu2000}. 
The condensation of such particle in the ground state can be described by classical scalar fields, of which the time evolution is governed by Schr\"{o}dinger-Possion equation in the non-relativistic limit.


Under the assumptions that DM can be considered as superfluid to describe the dynamics of the scalar field, a quantum-mechanical stress tensor acts as effective pressure that counters gravity and resists compression below a characteristic Jeans scale $k_J$ \citep[see, e.g.][]{Woo2009,Hui2017,Zhang2018}. 
Although often dubbed as "quantum pressure", this "pressure" term originates from basic quantum principles without any relations to classical thermodynamics \citep{Nori2018}. 
With typical bosonic mass of the order of $m \sim 10^{-22}$ eV, FDM scenario predicts flat, core-like density profiles within 1 kpc of the center of halos below $10^{10} M_{\odot}$ at z = 0,  providing a possible solution to the cusp-core problem \citep[see,e.g.][]{Hu2000,Marsh2014,Schive2014,Martino2018}.

The $k_J$ in FDM evolves slowly with time in matter-dominated epoch ($k_J \propto a^{1/4}$), which leads to a sharp cutoff in the matter power spectrum. 
Therefore the FDM model suppresses the small-scale galaxy and halo abundance while it also inherits the successes of CDM on large-scale structure \citep[see,e.g.][]{Woo2009,Marsh2010,Schive2015}. 
Moreover, the steepness of the small-scale break in FDM power spectrum may alleviate the {\it Catch 22} problem that WDM models have encountered \citep{Hu2000, Marsh2014}; i.e. WDM cannot be tuned to produce the required suppression on small-scale structure while simultaneously providing large enough cores suggested by observations \citep{Maccio2012}, although the window of mass range might be small \citep{Marsh&Pop2015}.

Given that the differences between CDM and FDM models present at small scales, constraints on FDM models are often obtained from two regimes. 
Studies of kinematic properties of Milky Way or Local Group dwarf galaxies attempt to fit the solitonic core structure in the FDM scenario.
Such studies, based on different analysis procedures and observational data sets, favour the FDM mass with $m \sim 1-6 \times 10^{-22}$eV \citep{Schive_Chiueh2014,Marsh&Pop2015,Calabrese2016,Gonzalez2017,Martino2018}.

A completely different regime to constrain FDM models comes from studying its impact on structure formation at high redshift.
Key to this approach is the fact that the small-scale systems form the earliest in the standard hierarchical structure formation. Thus at high redshift, the differences in structure formation between CDM and FDM scenarios are rather dramatic.
The largest differences are expected at epochs prior to when the small-scale formation starts to get entangled with the evolution of large-scale matter distribution; at later times, the discrepancies get smaller as structure formation becomes increasingly non-linear.

In this regime, the Lyman-$\alpha$ forest is a powerful probe to the DM distribution at the small scale at $z = 2\sim5$.
Recent studies utilize the Lyman-$\alpha$ forest flux power spectrum to give tight constraints on FDM bosonic mass of $m \gtrsim 2 \times 10^{-21}$ eV with 95\% confidence level \citep{Armengaud2017, Irsic2017, Kobayashi2017}, which is much higher than the values typically favoured by studies of nearby galaxy cores. 
The tension between Lyman $\alpha$ and the cusp-core constraints may be alleviated by accounting for the underlying systematic uncertainties, for example, assumptions on gas temperature and inhomogeneous re-ionization history \citep[see, e.g.][for more discussions]{Zhang2017, Hui2017}, such that the lower bound of Lyman-$\alpha$ forest constraints might loosen to a few times $10^{-22}$ eV \citep{Zhang2017}.

Another powerful constraint comes from observations of faint-end galaxies at higher redshifts.
Rapid progress has been made in probing $z \geq 6$ faint-end galaxy UV luminosity functions (LFs) in the past decade, thanks to deep surveys carried out by Hubble Space Telescope (HST) and applications of techniques such as lensing magnification by galaxy clusters (e.g. the Hubble Frontier Field (HFF) program). 
A number of works \citep[e.g.][]{Bozek2015,Schive2015,Menci2017,Carucci2018} have already discussed how the faint-end LFs can be sensitive to small-scale structure formations and provide a powerful tool to constrain FDM models.
In particular,  semi-analytic models or DM-only cosmological simulations have been performed to quantify the suppression of the small halo abundance in FDM scenario and then been used to predict the high-redshift galaxy UV LF (using empirical relations between galaxy luminosity and halo mass). 
Using this approach, \cite{Schive2015} constraints FDM with bosonic mass of $m \gtrsim 1.2 \times 10^{-22}$ eV (2$\sigma$ confidence level).  
Based on \cite{Schive2015} simulations, \cite{Menci2017} extended the constraint to $m \gtrsim 8 \times 10^{-22}$ eV ($3\sigma$) by deriving cumulative galaxy number densities and adding the more recent LF measurement from the HFF \citep{Livermore2017}. 
However, there are yet no hydrodynamic simulations that directly model galaxy formation in FDM scenario to compare with observations of galaxy populations and faint-end UV LFs at high redshift.

In this study, we perform full hydrodynamical cosmological simulations to accomplish this goal. Building upon the success of the \textsc{BlueTides} simulation, which has unprecedented large volume (400 Mpc/$h$ side box) to provide statistical validation against current constraints of high-z galaxy LFs down to $z \sim 7$ \citep{feng2016}, we use the same simulation code and tested implementation for the galaxy formation physics in this study with the additional FDM structure formation scenarios. We focus on FDM mass in the range of $m_{22} = 1 - 10$, where $m_{22} \equiv m/10^{-22}$ eV, and utilize recent results of high-redshift galaxy LF observations to place constraints on the FDM mass. 
We also discuss the potential impact of sub-grid physics and cosmic variance on the FDM mass limits, which are usually ignored in previous studies (and we shall show can lead to more optimistic constraints for FDM).

The paper is structured as follows. 
In Section~\ref{section2:Methods} we introduce our method, including a brief summary of the sub-grid models applied in our hydrodynamic code and how we implement the FDM physics in the initial condition setup. 
In Section~\ref{section3:Results} we present our results on the galaxy stellar mass functions and UV LFs, and compare them with current observations to derive limits on FDM mass. 
Finally, we summarize our results in Section~\ref{section4: Conclusion}.


\section{Methods}
\label{section2:Methods}

\begin{figure}	
\includegraphics[width=0.95\columnwidth]{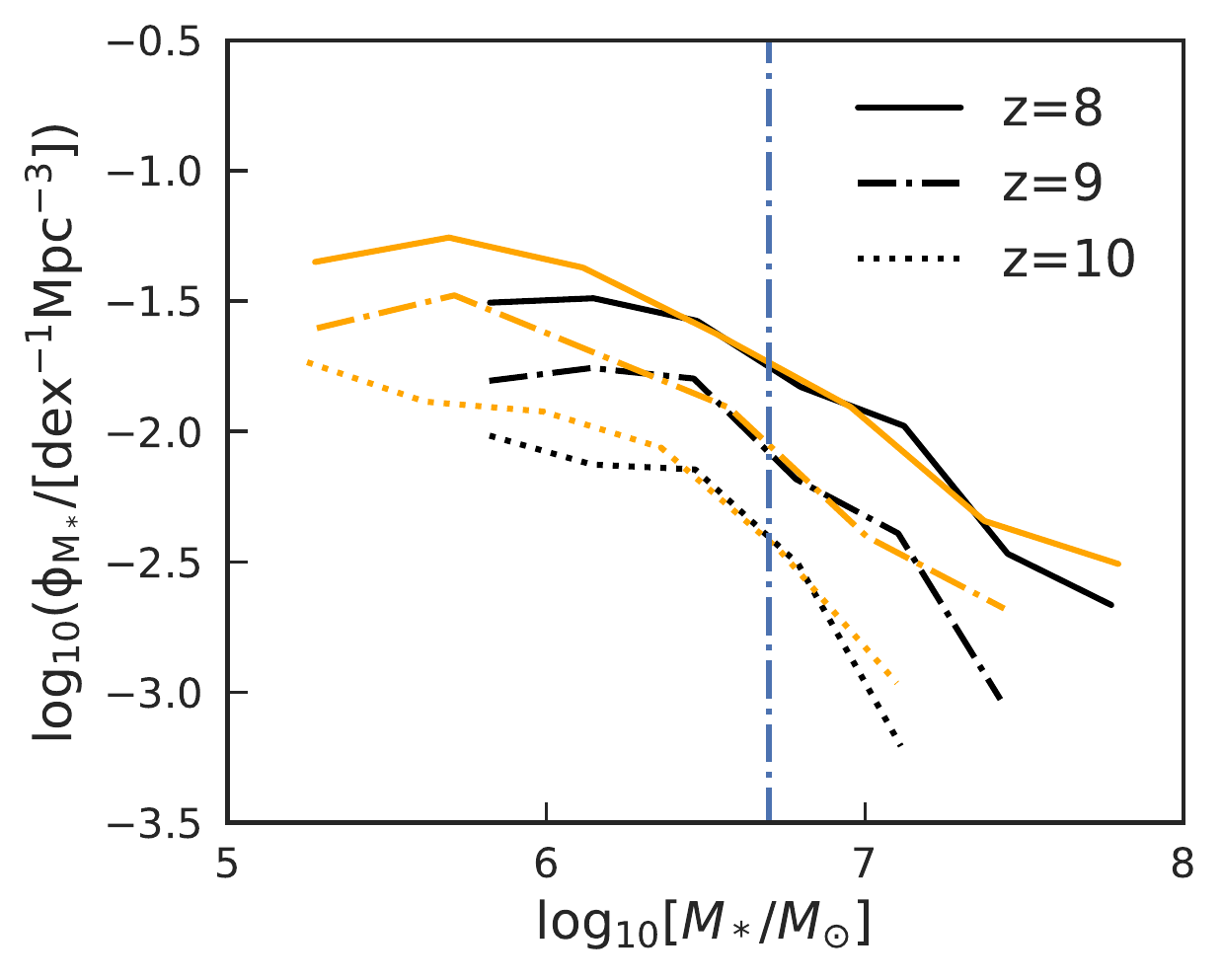}
\caption{
GSMFs from $z=8-10$ for CDM runs with $N=2\times 512^3$ (orange lines) and $N=2\times 324^3$ (black lines) to verify the convergence for the predicted galaxy abundance (at $M_* > 5\times 10^6 M_\odot$, which is shown by the vertical blue dash-dotted line).}
\label{fig:rsl}
\end{figure}

\begin{figure}
\hspace{-0.2cm}
	\includegraphics[width=1\columnwidth]{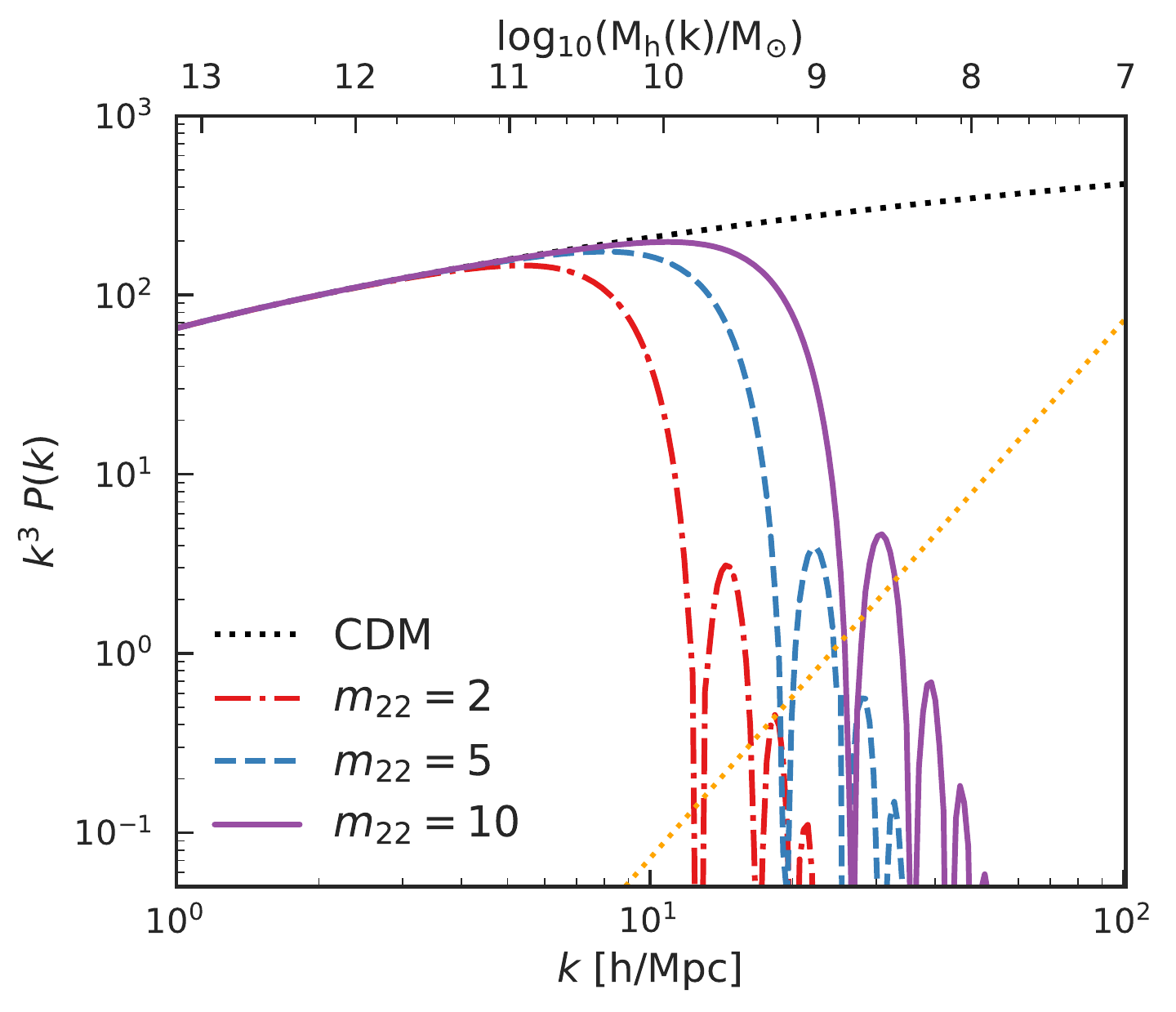}
    \caption{Dimensionless linear matter matter power spectra at $z=0$ for CDM and three FDM models of $m_{22}$ = 2, 5, and 10. The FDM power spectra are obtained by AxionCAMB code. 
    In the upper $x$-axis we convert the wavenumber to halo virial mass using $\rm M_h = 4 \pi (\pi/k)^3 \rho_m/3$, where $\rm \rho_m$ is the comoving matter density.
    The orange dot line shows the shot noise component in the simulations.
    }
    \label{fig:figure1}
\end{figure}

\subsection{Simulation setup}
\label{subsection:SimulationSetUp}
We perform hydrodynamical simulations within the CDM and FDM models using the Smoothed Particle Hydrodynamic (SPH) code \textsc{MP-Gadget} \citep{feng2016} with a well-tested implementation of galaxy formation models \citep{feng2016, DiMatteo2017, Wilkins2017, Wilkins2018, Bhowmick2018}. Here we briefly list some of its basic features, and we refer the readers to those papers for detailed descriptions and associated validation for a number of observables in the high$-z$ regime. 
In the simulations, gas is allowed to cool through both radiative processes~\citep{Katz} and metal cooling. 
We approximate the metal cooling rate by scaling a solar metallicity template according to the metallicity of gas particles, following the method in \cite{Vogelsberger2014}.
Star formation (SF) is based on a multi-phase SF model ~\citep{SH03} with modifications following~\cite{Vogelsberger2013}.
We model the formation of molecular hydrogen and its effects on SF at low metallicities according to the prescription by \cite{Krumholtz}. 
We self-consistently estimate the fraction of molecular hydrogen gas from the baryon column density, which in turn couples the density gradient into the SF rate.
We apply type II supernova wind feedback model from \cite{Okamoto}, assuming wind speeds proportional to the local one dimensional DM velocity dispersion. 
A difference in our simulations here is that (unlike \textsc{BlueTides}), we do not incorporate a patchy reionization model.
Instead, we use homogeneous UV radiation background from UVB synthesis model provided by \cite{fg09} that encapsulates a set of photoionization and photonheating rates that evolve with redshift for each relevant ion.

Cosmological parameters in the simulation are set as follows:  mass density $\Omega _{\rm m} = 0.2814$, dark energy density $\Omega _{\Lambda} = 0.7186$, baryon density $\Omega _{\rm b} = 0.0464$, power spectrum normalization $\sigma _{8} = 0.820$, power spectrum spectral index $\eta _{s} = 0.971$, and Hubble parameter $h = 0.697$. 
The main suite of simulations are generated with box size of $(15 h^{-1} \rm Mpc)^3$ periodic volume with $2 \times 324^3$ particles. 
Force softening length is 1.54 $h^{-1}$ kpc for both DM and gas particles,
and the corresponding mass resolution is $6.5 \times 10^6 h^{-1} \rm M_\odot$ for DM particles and $1.28 \times 10^6 h^{-1} \rm M_\odot$ for gas particles. 
An additional CDM high-resolution run with $2 \times 512^3$ particles is generated to test the numerical convergence. Its initial condition is set to be the same as its low-resolution counterpart. 
All of our simulations, except the high-resolution run, evolve from $z = 99$ to $z = 5$, with the initial conditions of CDM and three FDM models sharing the same random field but with different input matter power spectra. 
The features of FDM linear matter power spectrum will be described in the next section.
We run several FDM simulations with bosonic mass of $m_{22}$=1.6, 2, 5, and 10. This choice of mass range covers the values that are relevant for generating solitonic structure in dwarf galaxies ($m_{22} \sim 1$) \citep{Schive2014} up 
to current constraints inferred from the Lyman-alpha forest studies ($m_{22} \sim 10$) \citep{Armengaud2017}.

Following previous work which used N-body simulations to study high-redshift structure formation in the FDM framework \cite[e.g., ][]{Schive2015,Armengaud2017,Irsic2017}, we do not encode the quantum pressure (QP) term in the dynamic of our FDM simulations.
This should not alter our predictions significantly because for the FDM mass ($m_{22} \gtrsim$ 2) and the redshift range ($z \gtrsim 6$) considered in our simulations, the suppression in initial power spectrum is the dominant contributor to the suppression in structure formation. 
This has been studied directly in recent simulations by \cite{Zhang2017} and \cite{Nori2019} where the full treatment of FDM included the QP term.
This work (DM-only simulations) was able to show that encoding the QP does provide an extra but small suppression on halo abundance around $M_{\rm h} = 10^{10} M_\odot$ (see discussion in Section~\ref{section4: Conclusion}).
This level of accuracy in the predictions is beyond the scope of this work. 
In future work we will consider this effect as it may be required when future observations (e.g. with JWST) that are expected to provide tighter constraints on the faint-end galaxy abundance will become available. 

To compare our simulation results with the high-redshift observations of UV LFs, we build spectral energy distributions (SEDs) for each star particle in the simulated galaxies which provide mass, SFR, age and metallicity. Specifically we use the Binary Population and Spectral Populations Synthesis \citep[BPASS,][]{Eldridge} models with a modified Salpeter initial mass function (IMF) with a high mass cutoff at 300 ${\rm M_\odot}$.
In earlier work \citep{Wilkins2017} this method was to show that the high-redshift galaxy luminosity in the \textsc{BlueTides} simulation matches well with current observation constraints.
\yueying{
Note we have compared the luminosity constructed from the full SED modelling with a method that utilize a simple scaling relation with respect to star formation rate, $M_{\rm UV} = -2.5 \log_{10}(\dot{M_*})-18.45$ \citep[see, e.g.,][]{Stringer2011}, where $\dot{M_*}$ is the star formation  rate in unit of $M_\odot$/yr. 
We find that this simple scaling relation reproduces results of our SED models very well.
}
\yueying{
UV observations may be affected by dust attenuation, but this is only important for bright-end of the LFs ($M_{\rm UV} \lesssim -20$) \citep[see,][]{Wilkins2017}.
Since we only focus on the faint-end of the LFs where the discrepancies between CDM and FDM model predictions occur (${\rm M_{UV}} \gtrsim$ -17), we therefore do not make an attempt to include dust as its effects are negligible due to the relatively low stellar mass and metallicities in those galaxies with $M_* \lesssim 10^9 M_\odot$.
}

Our choice of box size of $(15 h^{-1} \rm Mpc)^3$ (similar to those
of \cite{Schive2015}) is adequate for deriving abundance of small mass halos and faint galaxies at high redshift, and is also comparable to the current effective volume of LF measurements through lensing magnification of galaxy clusters \citep[see discussion in][]{Livermore2017}. 
Surveys of small volume suffer from cosmic variance due to the underlying large-scale density fluctuation. This is an intrinsic limitation for distinguishing between different DM models.
Here, we estimate the cosmic variance, by taking advantage of the large volume \textsc{BlueTides} simulation (with box size of 400 Mpc/$h$ per side), and construct about $10^5$ subvolumes of $(15 h^{-1} \rm Mpc)^3$ box (comparable to HFF effective volume) and $(24 h^{-1} \rm Mpc)^3$ box (comparable to HUDF effective volume) in which we directly measure the expected cosmic variance on galaxy stellar mass function (GSMF) and LFs in the mass/luminosity range considered.

To investigate down to what mass scales we can expect to predict reliable galaxy abundances, we run a higher resolution simulation with $2\times 512^3$ particles.
While it is possible to run the main suite of simulations with higher resolution, it may not be completely appropriate. 
Changing the resolution significantly away from that employed in \textsc{BlueTides} would require a complete re-calibration of all sub-grid physics as we know that those have converged at resolutions similar to what we use here (this is by construction matched closely to that has been well-tested by \textsc{BlueTides}). 
Also as we will show later, the current resolution is adequate for estimating FDM model constraints using current observation data. 
To derive an appropriate resolution limit, we compare the constructed GSMF between our standard resolution runs and the high-resolution run, which are shown in Fig.~\ref{fig:rsl} for GSMF from $z=8-10$.
The two simulation results show good convergence for GSMF above $M_* \geq 5 \times 10^6 M_\odot$
(shown by the vertical dash-dotted line). Below this value, the galaxy abundance is under-estimated in the lower resolution run. 
Therefore we choose this stellar mass as our minimum galaxy stellar mass for the predictions throughout the rest of the paper.



\begin{figure*}
\hbox{
\hspace{-1.9cm}
\vbox{
\includegraphics[width=1.9\columnwidth]{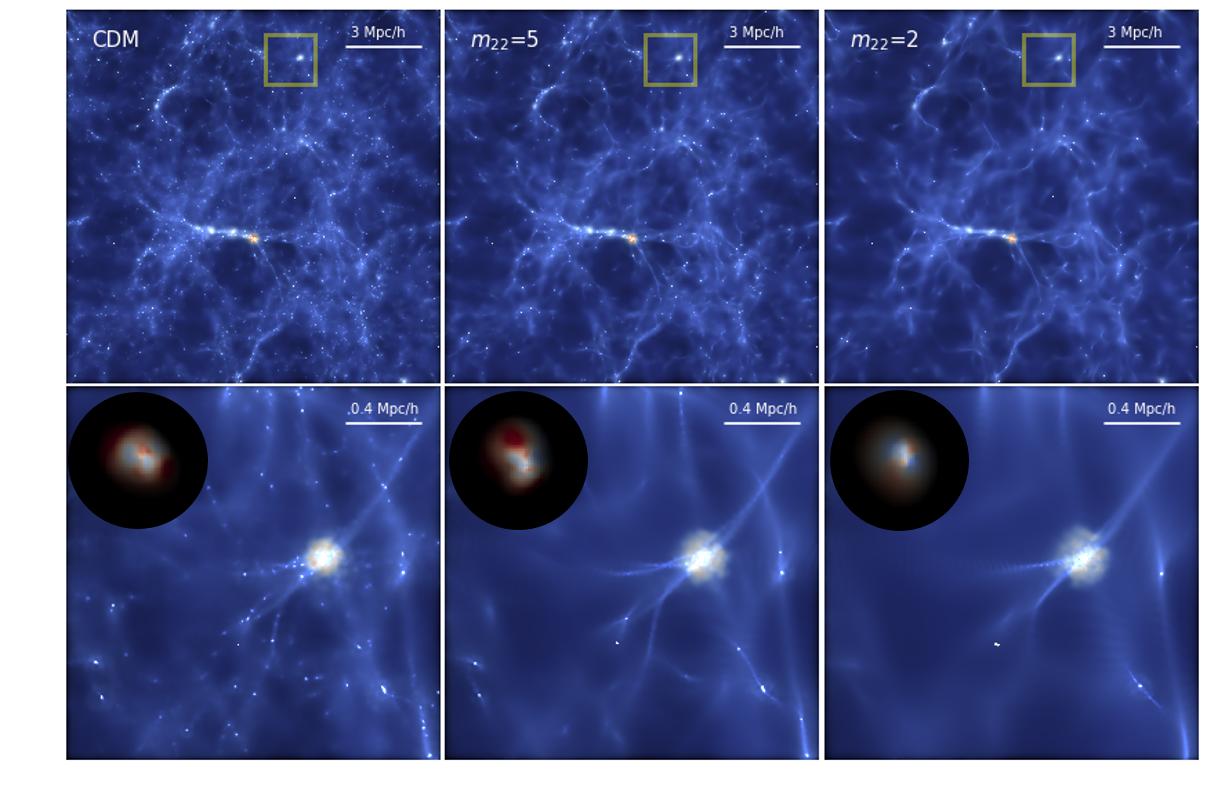}
}
\begin{centering}
\hspace{-1.4cm}
\begin{subfigure}[c]{0.2\textwidth}
\vspace{-10cm}
\includegraphics[width=0.8\columnwidth]{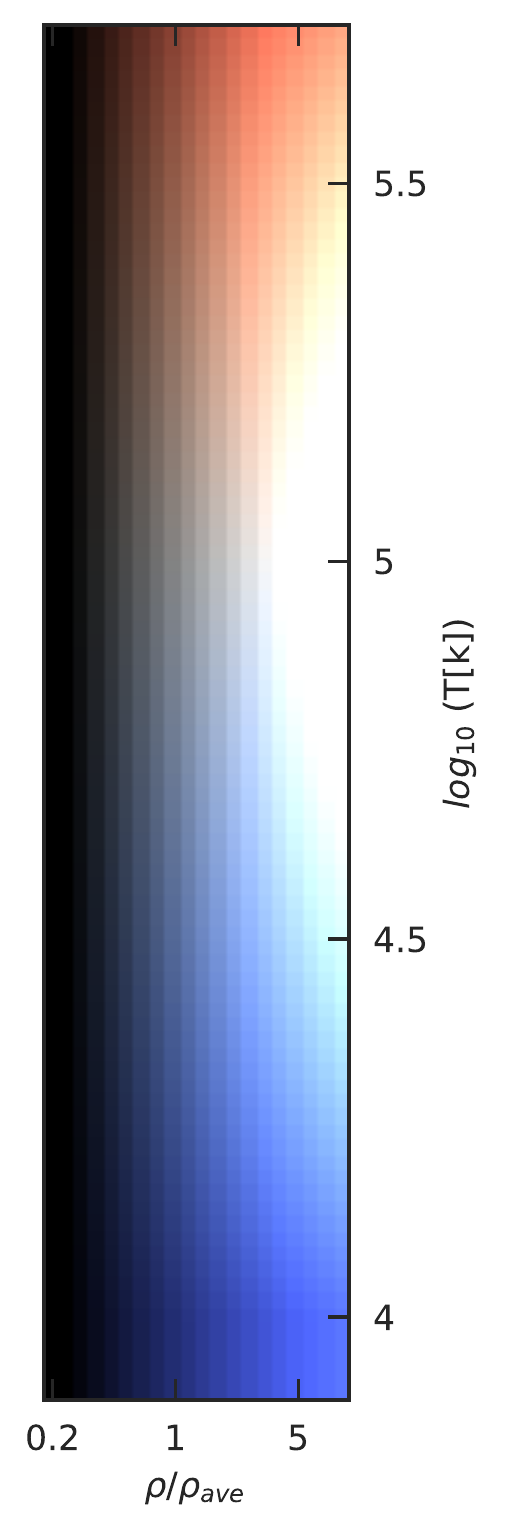}
\end{subfigure}
\end{centering}
}
\caption{
\textit{Upper panels} : Large-scale gas density field at $z=6$ in the simulations of CDM (left panels) and FDM models with $m_{22}$ = 5 (center) and $m_{22}$=2 (right). 
\textit{Bottom panels} : The zoomed-in regions of 2 Mpc/$h$ on the side (marked by the rectangular areas in the upper panels) to better illustrate the small-scale structures, in particular how they are smeared out in the FDM scenarios.
The gas density field is color-coded by temperature (blue to red indicating cold to hot respectively, see the 2D color bar aside).
The circular inserts in the bottom panels show the zoomed-in stellar density of the galaxy at the center of the panels. It is color-coded by the age of the stars (from blue to red, indicating young to old populations respectively). The radius of the zoomed-in circular region is 40 kpc/h and the mass of the galaxy is about $7 \times 10^8 \rm M_\odot$.
\textit{Rightmost panel}: The 2D color bar shows the values of the gas density and associated temperature. The temperature and density range are the same for all six panels. Here $\rho_{\rm ave}$ represents the average surface density shown in the volume. 
}
\label{fig:image}
\end{figure*}

\begin{figure}
\hspace{0.3cm}
	\includegraphics[width=0.9\columnwidth]{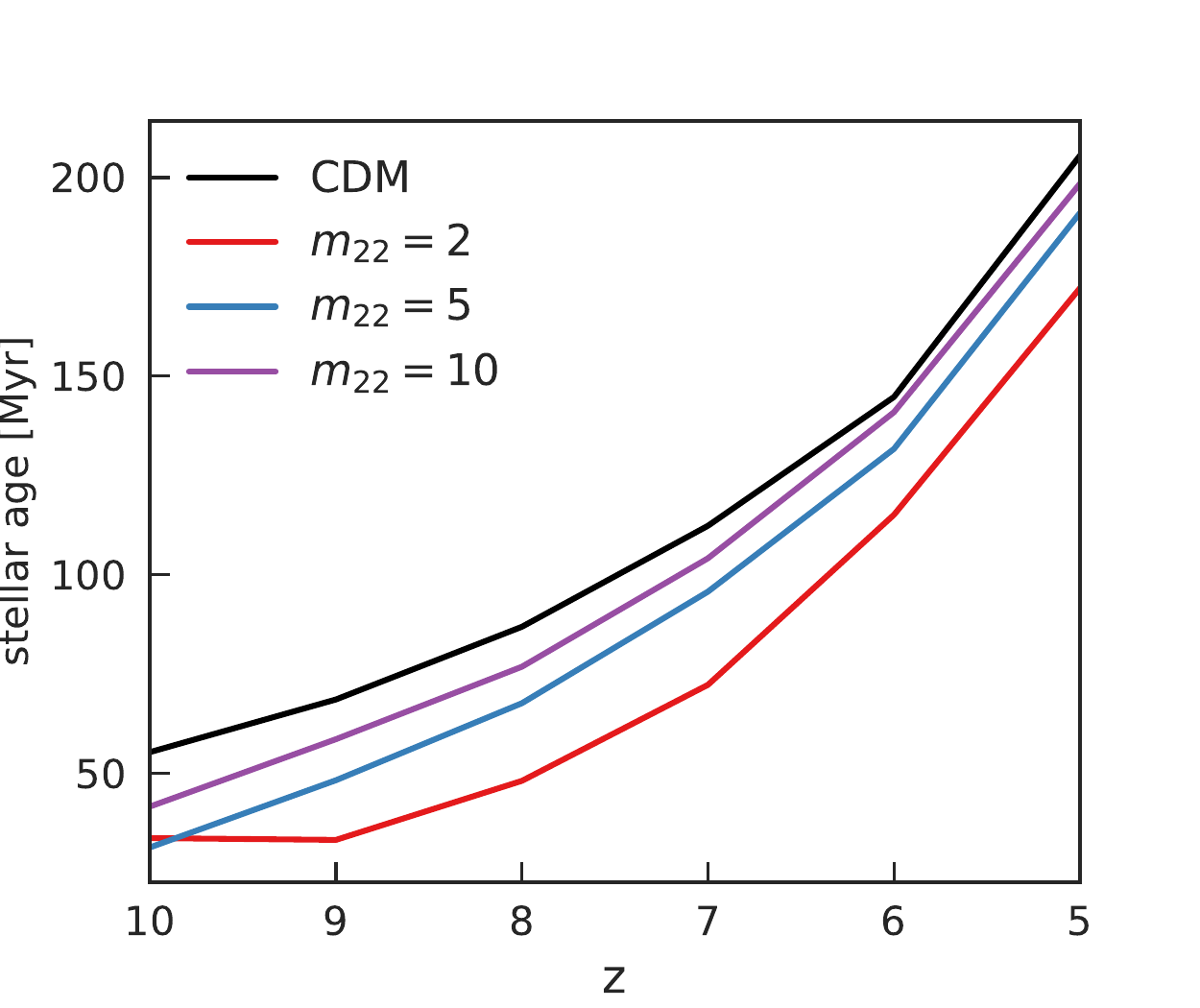}
    \caption{The globally average stellar age in CDM and the three FDM models as a function of redshift. The star formation process is systematically delayed in the FDM models compared to their CDM counterpart.}
    \label{fig:age}
\end{figure}

\subsection{Initial power spectrum}
\label{subsection: ICs}
The evolution of the FDM is governed by the Schr${\ddot{\rm o}}$dinger-Poisson equation which describes the dynamics of bosonic field. The balance between gravity and quantum pressure introduces a Jeans scale at $k_J \approx 69.1 m_{22}^{1/2}{\Omega_m h^2 \over 0.14} (1+z)^{-1/4} {\rm Mpc^{-1}}$.
For scales larger than $k_J$, i.e, $k < k_J(a)$, gravity dominates and the perturbation grows with time in the same way as CDM. For $k > k_J(a)$, the quantum pressure acts against gravity and the amplitude of perturbation oscillates with time.
The cutoff scale in the matter power spectrum compared to CDM can be characterised by the value of $k_J$ at the epoch ofmatter-radiation equality $k_{\rm J,eq} = 9 m_{22}^{1/2} \rm Mpc^{-1}$. 
Since $k_J$ grows with time, a small scale mode which was originally oscillating will eventually start to grow. Thus we can see a decaying oscillating feature in FDM power spectrum for $k \gtrsim k_{\rm J,eq}$.

Our FDM simulations employ the initial power spectrum at $z=99$ generated by \textsc{AxionCAMB} \citep{Hlozek}.  
Fig.~\ref{fig:figure1} shows the linear matter power spectra of the FDM models considered in this study at $z = 0$, with the CDM power spectrum plotted in the black dotted line.
For $m_{22}$ = 2, 5, 10, the corresponding $k_{\rm J,eq}$ is equal to 18, 28, 40 $\rm Mpc^{-1}$ $h$ respectively, consistent with the suppression scale seen on the corresponding linear power spectrum. In the upper $x$ axis of Fig.~\ref{fig:figure1}, we also show the typical halo mass for a particular matter power spectrum wavenumber. This allows us to get a better idea of what halo mass range the different FDM cutoffs scales affect. 
We convert the wavenumber to the halo virial mass using $M_h = 4 \pi (\pi/k)^3 \rho_m/3$, where $\rho_m = 3H_{0}^2\Omega_m/8\pi G$ is the background matter density of the universe.

We note that, in comparison with WDM models, the suppression in the FDM models with e.g. $m_{22}=5$ is similar to that of a WDM candidate of sterile neutrino with $m_\nu 
\sim 1.6$ keV through half-mode matching \citep[e.g. see derivations in][]{Marsh2016}. In that case, the power is suppressed by 50 percent comparing to CDM at the same half-mode scale $k_{1/2}$ (for this case $k_{1/2} \sim$ 10 $\rm Mpc^{-1}$). 
However, the cutoff in the WDM power spectrum is much smoother than in FDM, implying a significant difference in power spectrum shape
between the two. This makes it hard to directly map FDM from WDM models.
Although some previous studies have utilized some fitting formula to capture power spectrum shapes of different DM models \citep[e.g.][]{Murgia2017}, typically a full simulation with explicit FDM IC is highly preferable to draw more robust conclusions on FDM \citep[see e.g.,][for more discussion]{Armengaud2017,Irsic2017,Menci2017}

We note that the Jeans mass $m_J$ at $z<99$ for $m_{22} \geqslant 2$ is $\lsim 2\times 10^8 {\rm M_{\odot}}$, and it decreases for larger $m_{22}$ and lower redshift. 
Although this mass scale is well above our simulation resolution, additional effects on structure formation introduced by encoding the effects of quantum pressure in the simulation dynamics, as we mentioned in previous sections, is expected to be $<10\%$. 
The effect of quantum pressure is not significant on halos mass scales near the detection limit of current observations, which turn out to be about $\rm 10^{10} M_{\odot}$ \citep[see discussions in ][]{Schive2015}. 
Also, as we will show later, most of our luminous halos have $M_h \gsim 5\times10^{9} M_{\odot}$. Lower mass halos have inefficient star formation and rarely populate galaxies. Therefore taking the approach of including just the suppression in the initial power spectrum but not encoding the quantum pressure term should have minor effects on the derived FDM constraints using high-redshift galaxy abundance in the mass scales we study here.

\section{Results}
\label{section3:Results}
The major goal of our study is to use fully hydrodynamical cosmological simulations with tested models of galaxy formation to examine the predictions for high-$z$ faint-end galaxy abundances. In this section we will show how properties of different components, such as gas, stars, and DM in galaxies, get to be modified
by FDM compared to CDM.
We will also discuss in detail our predicted GSMF and LFs, which are the main high$-z$ observables that allow us to place constraints on FDM models. 

We start by showing a series of maps illustrating images of different scales of the intergalactic medium (IGM) in Fig.~\ref{fig:image}. 
The images show the simulated gas density field at $z=6$ for CDM (left column panels) in two of the FDM models with bosonic mass of $m_{22}=5$ and $m_{22}=2$ (middle and right column panels respectively). 
We show maps for the entire simulation volume (upper panels) and also
a subregion of the volume (lower panels) that hosts a star-forming galaxy with stellar mass $\sim$ $10^9~M_{\odot}$ (marked by the rectangular boxes in the upper panels with side length of 2 Mpc/$h$). The stellar densities of the galaxy are shown in the circular insets.  
The gas density field is colored by temperature, with scales shown by the 2D color bar in Fig.~\ref{fig:image}.
The upper panels illustrate that while the large-scale structures (e.g., filaments, nodes, voids... etc) of the gas distribution stay virtually the same between the runs with different DM physics, the small-scale structures are smeared out in the FDM models. 
The smaller the bosonic DM mass is, the more severe the small-scale structures have been washed out. 
This becomes particularly evident in the lower sub-panels where we zoom into the environment around one small halo.
Around this particular halo, the gas density is more concentrated and the colors show that the gas has been heated more in the CDM model compared to its FDM counterparts. 
As discussed in previous works, this is due to the fact that in FDM models, the volume occupied by the halos is systematically larger as a consequence of the delayed dynamical collapse of the halo \citep[see, e.g.,][]{Nori2019}.
Finally, in the circular insets in the lower panels of Fig.~\ref{fig:image}, we show the stellar density of the zoom-in regions with radius 40 kpc/$h$ centered on the galaxy, colored by the age of stars with blue to red indicating young to old.
In the FDM scenario, the galaxy is systematically younger than its CDM counterpart.
The average age of stars in the galaxy is 114, 108 and 88 Myrs for CDM and FDM $m_{22}$ = 5, 2 scenarios respectively. 
Generally, the galaxy formation process at high redshift is found to be delayed in DM models with primordial power spectrum cutoff features \citep[e.g. for WDM, see][]{Bode2001}.
To illustrate this further in our simulations, 
in Fig.~\ref{fig:age} we show the average, global stellar age in the whole volume as a function of redshift.
The global stellar age in the FDM models is smaller than the CDM prediction for a given redshift, and the discrepancies get smaller with decreasing redshift.
This is another illustration of why one should look at high redshift to see prominent effects of FDM models on the galaxy population.


\begin{figure}
	\includegraphics[width=\columnwidth]{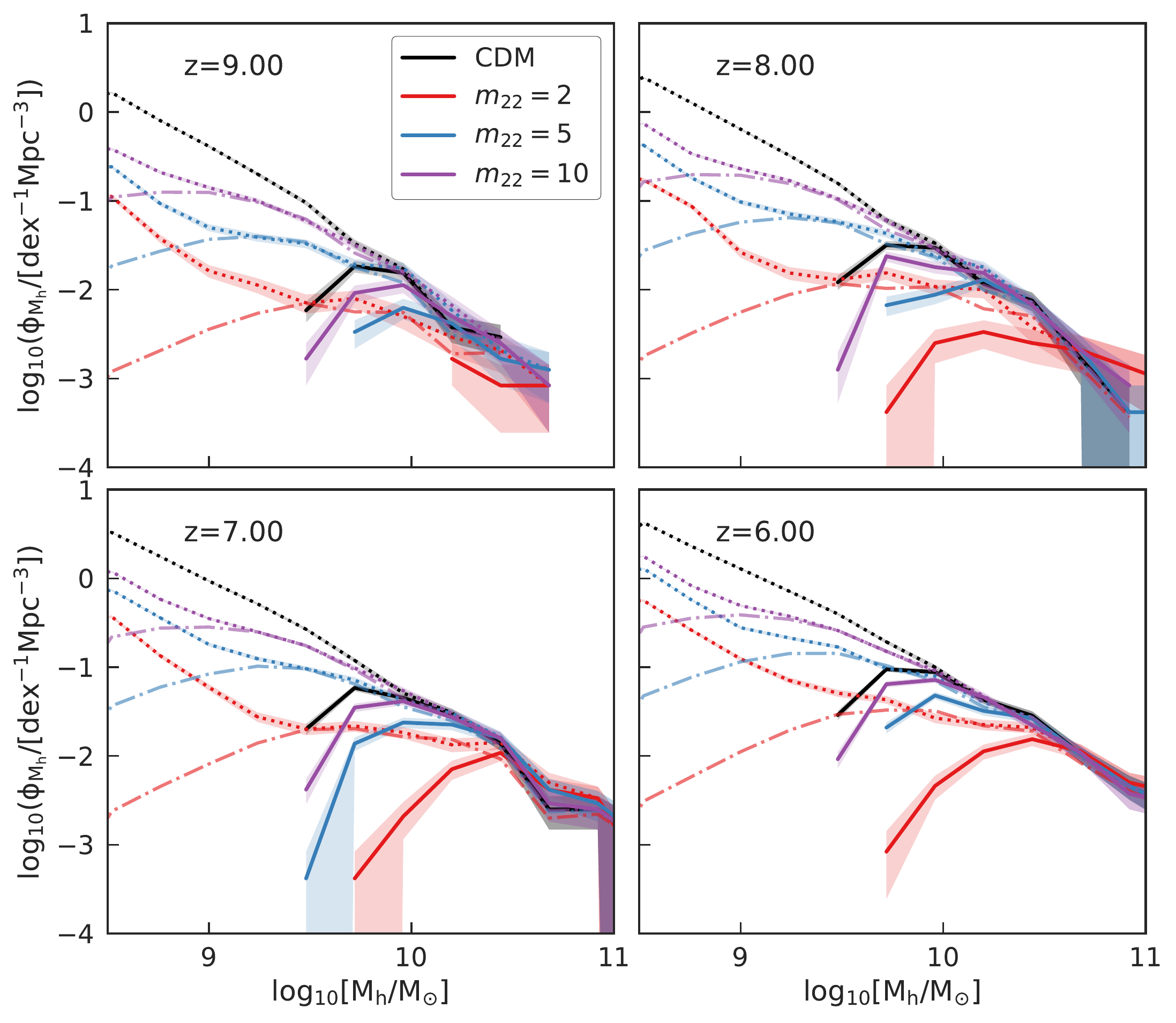}
    \caption{Halo MF of CDM and FDM models for $z = 6 - 9$. 
    Dotted lines are the original MFs constructed using the entire halo catalog.
    Solid lines represent the MFs of luminous halos, which are defined as halos that host galaxies with $M_* > 5 \times 10^6 M_\odot$.
    The shaded areas show the Poisson error bars. 
    Dash-dotted lines are the analytic halo MF fitting formula from \citet{Schive2015} (with spurious halos removed) for each FDM models.}
    \label{fig:hmf}
\end{figure}

\begin{figure}	
    \includegraphics[width=\columnwidth]{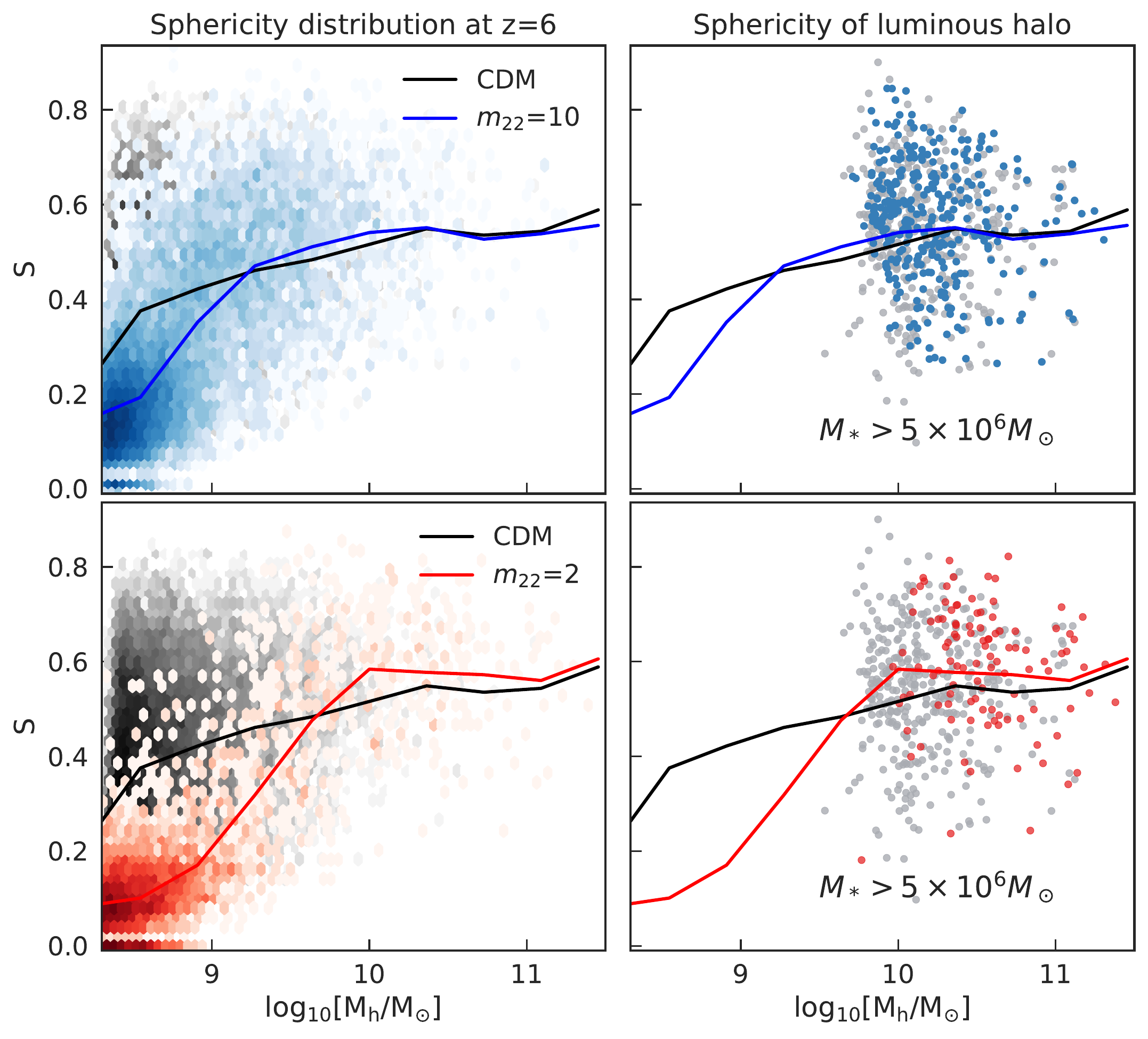}
    \caption{The comparison of halo sphericity distributions in CDM with those from two FDM models (top panels: with $m_{22}$=10; bottom panels: with $m_{22}$=2) at $z=6$. 
    The sphericity is calculated based on DM component only. The left panels show the distributions of all the halos in the two simulations, with grey contours representing CDM and colored contours for FDM models (blue: $m_{22}=10$; red : $m_{22}=2$). 
    The right panels show the population of luminous halos, with the same color scheme as the left panels. 
    Solid lines in all panels, which have the same color scheme as the contours, give the average sphericity of halos in the corresponding mass bins.}
    \label{fig:sp}
\end{figure}

\subsection{Impacts of FDM on the Halo Mass Functions}
\label{subsection:HMF}
To compare with previous work which employed 
N-body simulations of FDM, we start by showing the halo mass functions (MFs) for the four different DM models in the redshift range of $z=6-9$ in Fig.~\ref{fig:hmf}. 
The black lines show the CDM predictions whilst lines in purple, blue, and red are for FDM models with $m_{22}$ = 10, 5, 2 respectively.
The dotted lines are the original MFs constructed from the entire halo catalog.
Dash-dotted lines are the expected halo MF after the removal of spurious halos according to the analytic fitting formula provided by \cite{Schive2015}, as we will discuss later.
Solid lines are the halo MFs of luminous halos, and the shaded areas show their Poisson error bars. 

As expected from the input matter power spectra shown in Fig.~\ref{fig:figure1}, the halo MF is suppressed in FDM models compared to CDM.
\yueying{Take $z=6$ as an example, FDM halo abundance is suppressed by $\gtrsim50\%$ for halos in mass range $M_{\rm h} \lsim 10^{10} M_\odot$ for $m_{22}=2$ ($\lsim 5\times 10^{10} M_\odot$ for $m_{22}=10$).}
The suppression gets larger for smaller values of $m_{22}$ and becomes more prominent as redshift increases.
From the dotted lines in Fig.~\ref{fig:hmf}, we can see that at mass scales close to $M_{\rm h} \sim 10^{9} M_{\odot}$, an upturn feature in the halo MF, which is especially prominent for $m_{22} \leq 5$, is exhibited in FDM model results. This is likely due to a numerical artifact. 
Similar to WDM models that predict this small-scale structure suppression, it is known that FDM simulations also suffer from the problem of artificial fragmentation of filaments and contains spurious small halos \citep[see, e.g.][for WDM studies of spurious halos]{Wang2007,Lovell2014}. 
Several techniques have been proposed (as we will discuss later in this section) to detect and remove spurious halos for improving estimations of the true halo MF, upon which they can be used to infer the UV LF and compare with observations. 
However, we argue in the rest of this section that spurious halos have negligible effects on galaxy population because the luminous halos that host galaxies are more massive than the halo mass range where this numerical artifact becomes important.

To quantify the halo mass range for spurious fragmentation, \citet{Wang2007} provides an empirical estimation, $M_{\rm lim}=10.1 \bar{\rho}d/k_{\rm peak}^2$, of which below this halo mass scale $M_{\rm lim}$ halos are dominated by numerical artifacts rather than physical ones. Here $\bar{\rho}$ is the mean density of the universe, $k_{\rm peak}$ is the wavenumber at the maximum of the dimensionless matter power spectrum, and $d$ is the mean interparticle separation.  
In our FDM  model with $m_{22}=2$, $M_{\rm lim}$ $\sim 2\times 10^9 M_\odot$ (for $m_{22}=10$, $M_{\rm lim}$ $\sim 4\times 10^8 M_\odot$) corresponds to the upturn feature in our original halo MFs.

Other than simply applying a mass cut corresponding to $M_{\rm lim}$, additional discriminating criteria have been developed to refine the removal of those artifacts. One common method is to trace the Lagrangian region occupied by the halo member particles back to the simulation initial condition (protohalo) and use their spatial distributions (e.g., shape) as a proxy to discern the spurious ones. 
Another method is to match halos between different resolution runs for which a spurious halo would not be present in the higher resolution runs \citep[see, e.g.][for the discussions in the FDM scenarios]{Schive2015}. 
\cite{Schive2015} has provided an analytic fitting function for which we can derive the expected spurious-halo-free MF for a given FDM model. However, we note that this does not predict whether a halo is spurious or not, but rather shows the expected FDM halo population for a given halo mass range. In Fig.~\ref{fig:hmf} we show the "corrected-halo MF" as dash-dotted lines with spurious halos removed. 
Comparing this corrected-MF with MF of luminous halos (solid lines), we can see that the luminous halo population indeed doesn't extend to the halo mass region where the corrected-MF (dash-dotted lines) starts to deviate from original MF (dotted lines). Therefore it suggests that it is not necessary to apply spurious halo removal if we are concerned with deriving quantities from luminous galaxies in our simulations.    

As a further illustration, we also calculate the halo sphericity $S$ in the same way as defined in \cite{Schive2015}:
$S=\sqrt{\frac{I_1+I_2-I_3}{-I_1+I_2+I_3}}$, with $I_1 \leqslant I_2 \leqslant I_3$ the principle moments of inertia of the halos.
Note $S$ is calculated merely based on DM component. 
In Fig.~\ref{fig:sp} we show $S$ in CDM and two FDM models with $m_{22}=10$ (top panel) and $m_{22}=2$ (bottom panel) for halos identified at $z = 6$.  
It has been shown in previous studies that genuine halos are those with high $S$ (close to 1) while spurious halos occupy low $S$ regions. 
For example, \citet{Schive2015} suggests a cut of $S$>0.3 for removing spurious halos in the redshift range of $z = 4 - 10$. 
The solid lines in Fig.~\ref{fig:sp}, which show the average $S$ of each model as a function of halo mass $M_{\rm h}$, 
indicate that discrepancies of $S$ between CDM and FDM models are the most prominent at $M_{\rm h} \lsim 5 \times 10^9 M_{\odot}$ for $m_{22}$ = 2 and $M_{\rm h} \lsim 10^9 M_{\odot}$ for $m_{22}$ = 10. 
This is consistent with the $M_{\rm lim}$ we derived before.  
However, in the right panels of Fig.~\ref{fig:sp} where we only select the luminous haloes~(which will have the actual contribution to the calculated LFs), 
the distributions of $S$ of all DM models occupy the similar high $S$ regions with only a few exceptions.
All these results indicate that the luminous halos in our simulation are highly unlikely to be spurious halos.
This is due to the fact that very low mass halos do not have deep potentials to maintain enough gas to have efficient star formation. 
Therefore most of the low mass halos, including those generated by numerical artifacts, do not host galaxies. 
We hence conclude that the process of removing spurious halos is not required in our study of the high-$z$ galaxy population for which the effects remain negligible.



\begin{figure}
\hspace{0.5cm}
\includegraphics[width=0.9\columnwidth]{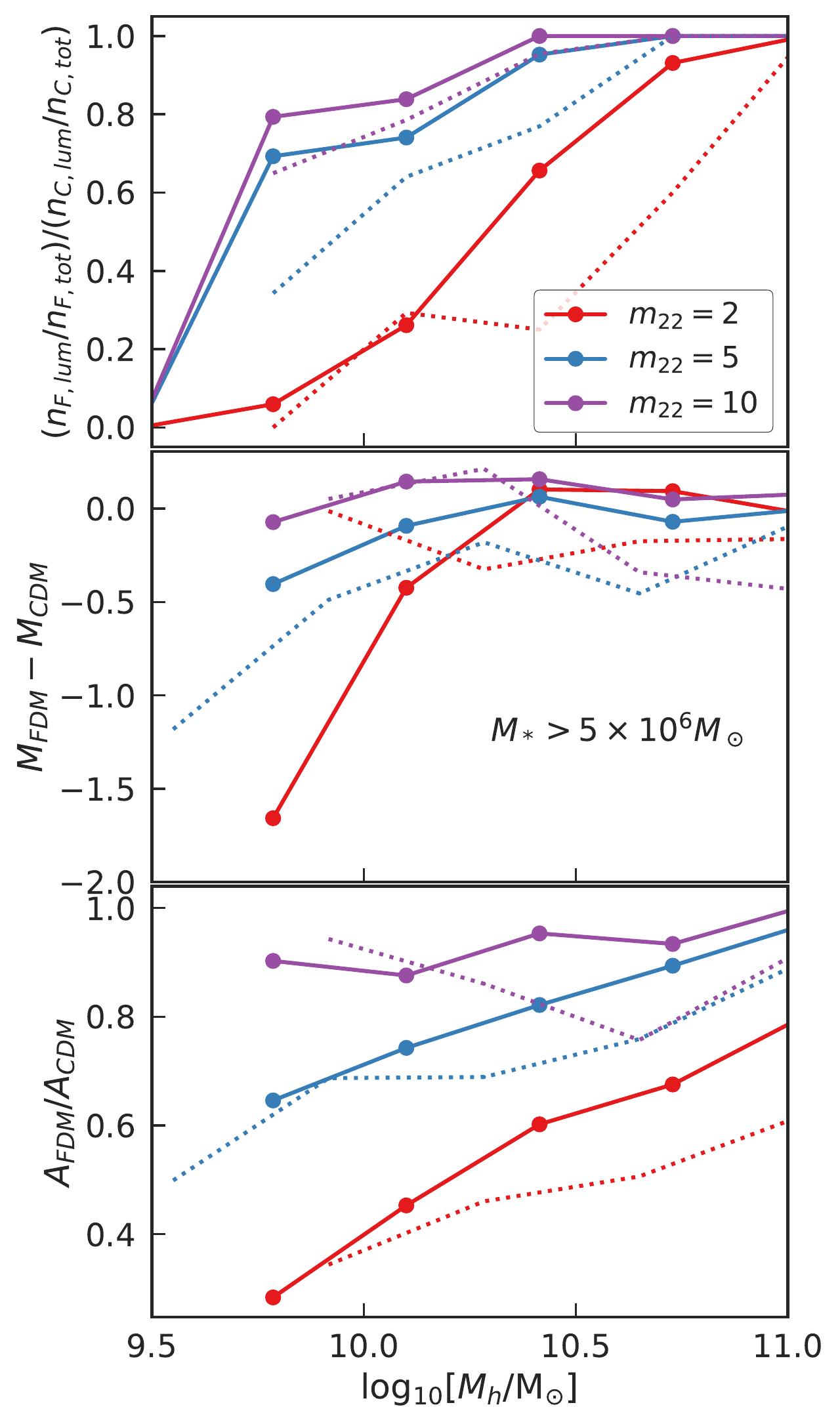}
\caption{
\textit{Top panel}: The ratio of the fraction of halos hosting luminous galaxies in FDM versus those in CDM as a function of halo mass (with $M_* \geq 5\times10^6 M_\odot$). 
\textit{Middle panel}: The difference in the UV band magnitude between FDM and CDM galaxies as function of halo mass.
\textit{Lower panel}: The ratio of average stellar age in galaxies between CDM and FDM models as a function of halo mass.
For all the three panels, solid lines show results at $z=6$, and the dotted lines are from $z=8$.
}
\label{fig:lumfrac}
\end{figure}

\begin{figure*}
    \includegraphics[width=2.2\columnwidth]{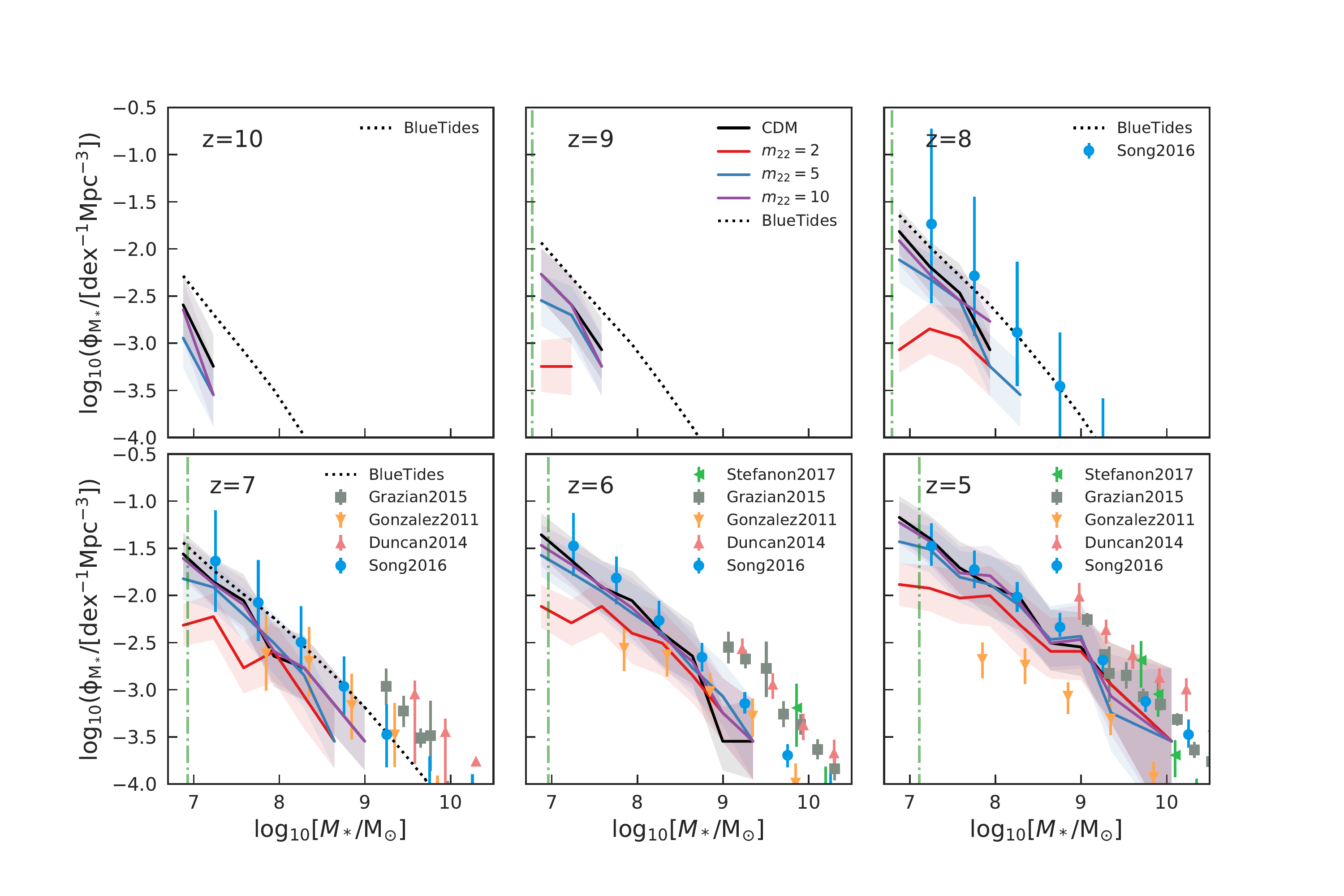}
    \caption{The GSMFs predicted in CDM (red lines) and FDM models with ${\rm m_{22}}$= 10, 5 and 2 (purple, green, and blue lines respectively) shown for $z= 5-10$.
    The color shaded areas are the 1 $\sigma$ cosmic variance uncertainties for a comoving volume of 15 Mpc/$h$ per side. This is estimated based on subvolumes drawn from the  \textsc{BlueTides} simulation as described in the text.
    The green vertical dash-dotted lines mark the limits for JWST deep field 
    Observational data points with 1 $\sigma$ error bars include results from \citet{Gonzalez2011} (orange triangles), \citet{Duncan2014} (pink triangles), \citet{Grazian2015} (grey squares), \citet{Song2016} (blue circles), and \citet{Stefanon2017} (green triangles). 
    }
    \label{fig:gsmf}
\end{figure*}

\begin{figure}	
\hspace{0.5cm}
\includegraphics[width=0.9\columnwidth]{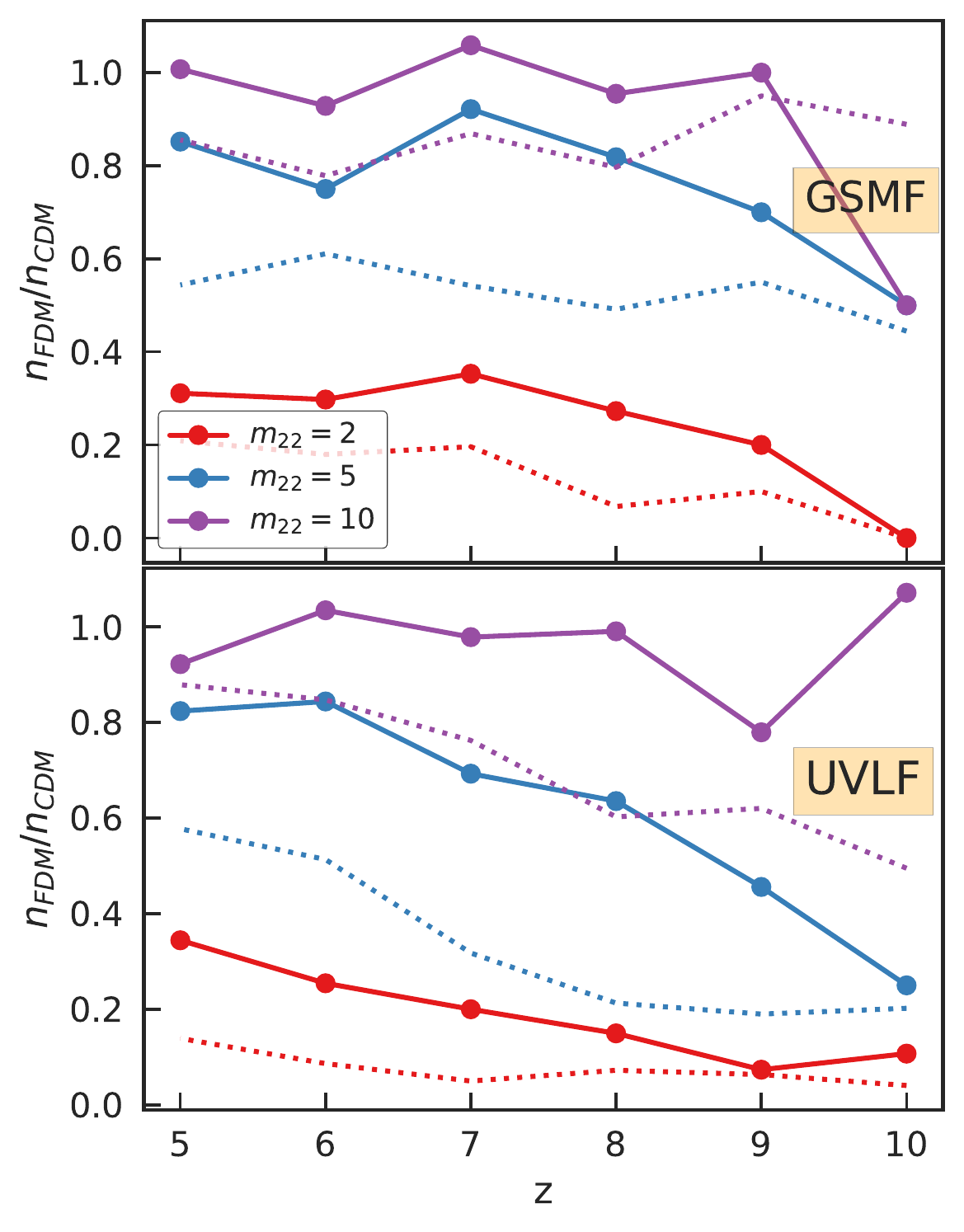}
\caption{
\textit{Upper panel}: The ratio of the luminous halo fraction in three FDM models versus CDM as function of redshift, which is derived from the GSMFs at two different stellar  mass bins: [$10^7 M_\odot$,$3\times10^7 M_\odot$] (solid lines), [$5\times 10^6$,$10^7 M_\odot$] (dotted lines).
\textit{Lower panel}: The same as the upper panel, but with UV LFs over two magnitude bands of [-14.5, -16] (solid lines) and [-13,-14.5] (dotted lines).}
\label{fig:z_evolution}
\end{figure}

\begin{figure}	
    \includegraphics[width=1.1\columnwidth]{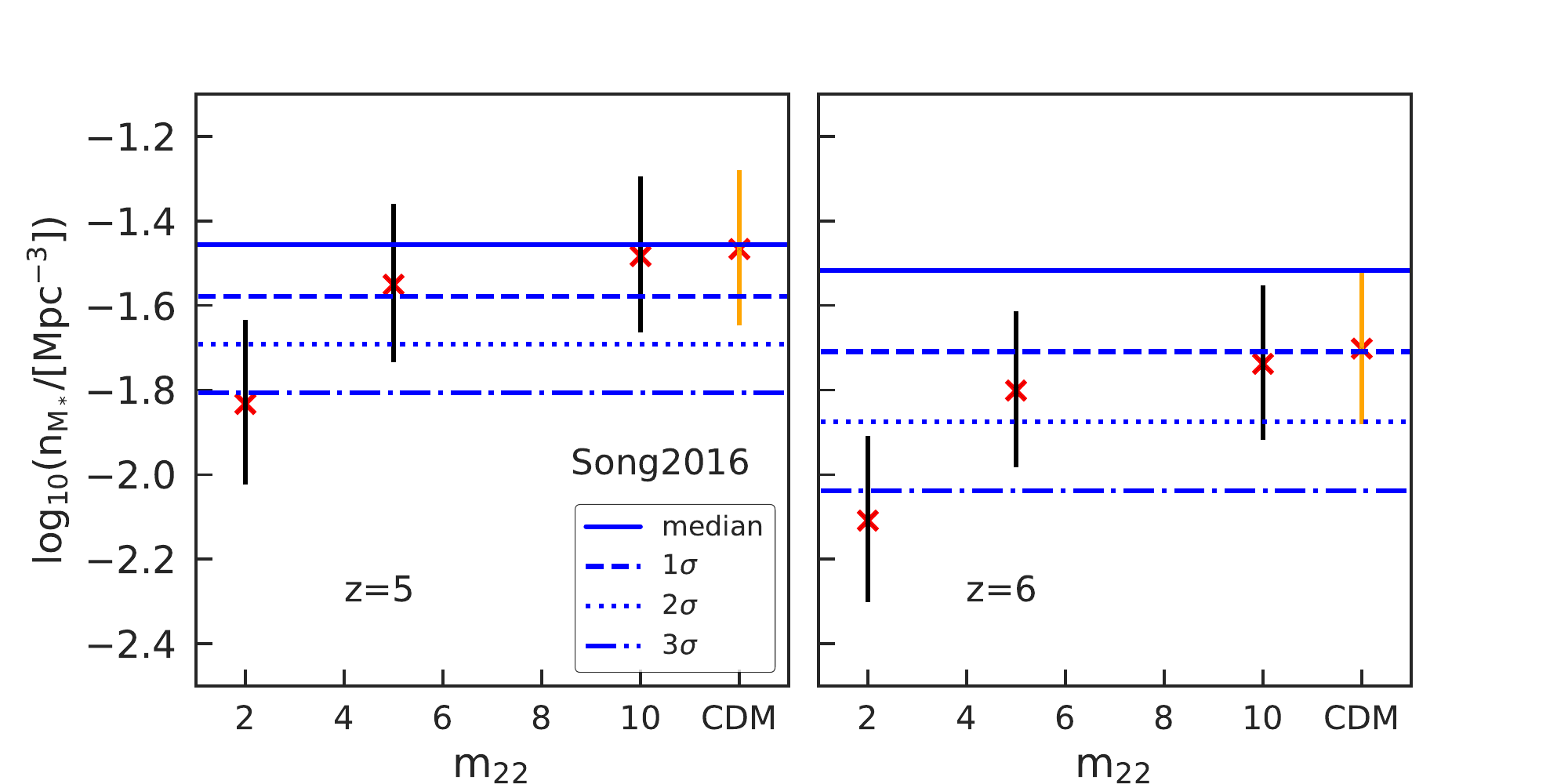}
    \caption{The cumulative number density of galaxies with mass $10^7 M_\odot < M_* < 10^9 M_\odot$ at $z=5$ (left panel) and $z=6$ (right panel) as function of $\rm m_{22}$. 
    The CDM results are marked by the rightmost orange points.
    We add $1\sigma$ cosmic variance on simulation data points based on volume of $\rm (15 Mpc/\it{h})^3$ and $\rm (24 Mpc/\it{h})^3$ (comparable to the HFF and HUDF effective volumes).
    The cosmic variance is estimated from the \textsc{BlueTides} simulation.
    The horizontal lines represent the observed galaxy number densities within a given mass bin with $1\sigma$ (dash lines), $2\sigma$ (dotted lines), and $3\sigma$ (dash-dotted lines) confidence levels. 
    The observational data set used here is from \citet{Song2016}.
    }
    \label{fig:nb_gsmf}
\end{figure}

\subsection{Galaxy Stellar Mass Function}
\label{subsection:GSMF}
Before we discuss the results for the GSMFs, we compare the population of luminous galaxies in CDM and FDM models as a function of halo mass. 
In the top panel of Fig.~\ref{fig:lumfrac}, we show the ratio of the luminous fraction
of halos hosting galaxies in FDM compared to CDM models as a function of halo mass. We define the luminous fraction, $n_{\rm lum}/n_{\rm tot}$ , as the fraction of halos that host at least one galaxy with $M_* > 5 \times 10^6 M_\odot$. 
The plot shows that below the halo mass, $M_h \lesssim 10^{10} M_\odot$, at which the FDM halo mass function starts to deviate from the CDM one the luminous fraction in FDM becomes significantly smaller ($\lesssim$ 70$\%$) than in CDM. 
More specifically, for example, in the FDM model with $m_{22}=2$  the luminous fraction is suppressed by $\sim 80\%$ for halos with $\rm M_h \sim 10^{10} M_\odot$ at $z = 6$. 
This indicates that the suppression on galaxy number density is caused not only by the decreased halo number density ($n_{\rm tot}$) in the FDM scenario but also by the fact that FDM halos are less likely to host galaxies compared to their CDM counterparts.

In the middle panel of Fig.~\ref{fig:lumfrac}, we show the difference in the UV magnitude for the galaxies (based on the stellar population synthesis model described in Section~\ref{subsection:SimulationSetUp}) in the FDM and CDM models as a function of halo mass. 
We find that for galaxies hosted by halos with $\rm M_h \sim 10^{10} M_\odot$, 
the average UV band luminosity at $z=6$ in FDM model with $m_{22}=2$ is about 0.5 mag smaller (brighter) than that in CDM.
Thus, we conclude that although haloes in the FDM model with $\rm M_h \lesssim 10^{10} M_\odot$ are less likely to host luminous galaxies compared to the ones in CDM; the galaxies that do emerge in FDM models are typically somewhat brighter than those in CDM.

The reason galaxies in FDM models tend to be more luminous than in CDM is likely due to the delayed star formation process in the former as this leads to a younger and brighter stellar population.
To directly examine the age of the stars in galaxies, we show the ratio of average stellar ages of galaxies between FDM and CDM as a function of halo mass (in the lower panel of Fig.~\ref{fig:lumfrac}). 
The galaxies in FDM models are indeed generally younger than those in CDM, an effect that becomes even more prominent as the halo mass decreases. 
For example, galaxies that reside in halos with $\rm M_h \sim 10^{10} M_\odot$ for FDM with $m_{22}=2$ at $z=6$ have average age about $40\%$ of their CDM counterparts, 
while for $\rm M_h \sim 3\times 10^{10} M_\odot$ (${\rm log_{10}M_h=10.5}$) the ratio is about 60$\%$.
This directly explains the difference in the galaxy UV magnitudes between CDM and FDM
(shown in the middle panel of Fig.~\ref{fig:lumfrac}).

Although the halo mass is a direct indicator of the underlying density fluctuation scale, it is not directly measured (at least in any of the high-redshift observations discussed in this work).
Next, we examine the direct predictions for the galaxy abundances as a function of stellar mass.
In particular, in Fig.~\ref{fig:gsmf} we show the GSMFs for CDM and FDM models in the redshift range of $z = 5-10$.
For comparison, the black dotted lines show the GSMFs in the \textsc{BlueTides} simulation, (which has only been run to $z=7$).
The shaded areas represent $1\sigma$ error from expected cosmic variance in a volume of 15 Mpc/$h$ side length.
Note that our smaller volume realization of CDM tends to be slightly lower than the one from the \textsc{BlueTides}, which is derived from a large simulation volume of 400 Mpc/$h$ side length. 
However, our results are well within the expected cosmic variance.

Fig.~\ref{fig:gsmf} shows that there is a suppression on galaxy abundance in FDM compared to CDM, and the suppression increases with decreasing stellar mass. 
For example, at $z=6$ the galaxy abundance is suppressed by $\gtrsim 50\%$ for stellar masses $M_* \lesssim 10^7 M_\odot$ $(\lesssim 10^8 M_{\odot})$ in the FDM model with $m_{22}=5$ ($m_{22} = 2$).
To quantify how the suppression evolves with redshift, we plot, in the upper panel of Fig.~\ref{fig:z_evolution}, the ratio of galaxy abundance in FDM over that in CDM (different line styles show different stellar mass range) as a function of redshift. 
This shows that the suppression of the galaxy abundance brought in by FDM  decreases for decreasing redshift. 
For example, the abundance of galaxies (from GSMF) with stellar mass range of $1-3 \times 10^7 M_{\odot}$ (solid lines) is suppressed by $\sim 40\%$ in the FDM model $m_{22} = 5$ (in blue) at $z=9$ , while this suppression is reduced to $\sim 20\%$ at $z=5$. 

Next, we compare the predictions for GSMFs from FDM models with the current observational constraints.
Observational measurements of the GSMF are typically obtained by taking the observed rest-frame UV LFs and convolving them with a stellar mass versus UV luminosity distribution at each redshift. 
The advantage of using GSMFs is that GSMFs can be easily derived from our hydrodynamic simulations, once run, without any further assumptions.
In Fig.~\ref{fig:gsmf} we show the current available observational data on GSMFs ($\rm \phi_{M_*}$) collected from \cite{Gonzalez2011} (orange triangles), \cite{Duncan2014} (pink triangles), \cite{Grazian2015} (grey squares),  \cite{Song2016} (blue circles), \cite{Stefanon2017} (green triangles).
Among these studies, \cite{Song2016} provides the strongest constraints on $\rm \phi_{M_*}$ for FDM as they probe out to the highest redshift ($z=8$) and to the smallest stellar masses ($M_* \lsim 10^7 M_\odot$). 
Their measurements are obtained through the combination of $\it HST$ imaging data together with the deep IRAC data from $\it Spitzer$ $\it Space$ $\it Telescope$ over the Great Observatories Origins Deep Survey (GOODS) fields and the HUDF.
In Fig.~\ref{fig:gsmf} we can see that, at $z = 5 - 8$, both our predictions for $\rm \phi_{M_*}$ in the CDM and FDM model with $m_{22} = 10, 5$ are consistent with the \cite{Song2016} data, while the $m_{22} =2$ model appears inconsistent for stellar masses below $M_* \lsim 10^8 M_{\odot}$. 
Note that, although the \cite{Gonzalez2011} data agrees with our $m_{22} =2$, it is systematically lower than the rest of the observational data sets. Therefore we do not take it into consider when we derive our FDM limits.  

We now wish to quantify more precisely the level of discrepancy between the model and observational measurements.
To do so we adopt a procedure similar to \cite{Menci2017}. 
In particular, we calculate the expected cumulative number density $\rm n_{obs}$ from data within the stellar mass range of $M_* =\rm 10^7 - 10^9 M_\odot$ and compare it with the abundance predicted from the simulations for each DM model.
To quantify the observational uncertainties, we rebuild $10^7$ realizations of galaxy stellar mass distributions by taking random value $\rm \phi_{M_*}$ in each mass bin according to a log-normal distribution with the variance given by the corresponding error bars.
Then we calculate the cumulative number density $\rm n_{obs}$ for each realization and derive the median and the uncertainties in $\rm n_{obs}$ from the constructed distribution (out of the $10^7$ $\rm n_{obs}$ realizations).
In Fig.~\ref{fig:nb_gsmf} we compare the $\rm n_{obs}$ derived from \cite{Song2016} with our simulation results at $z = 5$ and $z = 6$.
We choose these two redshifts because the observational data set has smaller error bars and thus should provide the most competitive constraint for FDM models.
The horizontal lines show the median (the solid lines) and $1\sigma$, $2\sigma$, $3\sigma$ lower bounds (the dashed, the dotted, and the dash-dotted lines respectively) of $\rm n_{obs}$ derived from the observations.
The data points marked with the red cross give the cumulative number density, $n_{M_*}$, obtained from our simulations for the different FDM models, while the CDM predictions are shown by the orange points (on the right). 
By comparing the lower bounds of $\rm n_{obs}$ with our predicted $n_{M_*}$, we infer that the FDM model with $m_{22}<2$ is ruled out at the $3\sigma$ confidence level by this data set. 
If we interpolate our models, we obtain that $m_{22}\lesssim 4$ can be ruled out by $2\sigma$ confidence level at $z=6$, although the constraint is slightly weaker at $z=5$.

The galaxies with inferred mass log$(M_*/M_\odot) < 8.5$ ($z=5-8$) primarily come from the HUDF observations (a 2.4' $\times$ 2.4' field) \citep{Song2016}. 
In Fig.~\ref{fig:nb_gsmf} we also add the cosmic variance on $n_{M_*}$ for volume of $\rm (24 Mpc/\it{h})^3$ which is comparable to HUDF survey volume.
As described in Section \ref{subsection:SimulationSetUp}, the expected cosmic variance is calculated directly from the galaxy population in the large volume \textsc{BlueTides} simulation.
Taking this into account, we infer that the predicted $n_{M_*}$ from the $m_{22}=5$ model is within the $1\sigma$ of the cosmic variance for the CDM model at both $z=5$ and $z=6$, indicating that in the mass band $M_* =\rm 10^7 - 10^9 M_\odot$, 
a survey of HUDF volume will not be able to distinguish clearly between $m_{22}=5$ and CDM (within cosmic variance). 
Surveys extending to small mass or with larger volumes may thus be needed to put further constraints on FDM models with $m_{22} \gtrsim 5$ from the GSMF.
\yueying{
In Fig.~\ref{fig:gsmf}, we show the expected detection limit for JWST deep field with survey volume comparable to HUDF in the vertical green dash-dotted lines. This is expected to probe to $M_* \lsim 10^7 M_\odot$.
The detection limit estimated for JWST lensed field with $10\times$ magnification can further extend down to $M_* \sim 10^6 M_\odot$ \citep{Yung2019_gsmf}.
To roughly estimate how JWST can put constraints at this stellar mass region, we use galaxies below our mass threshold of galaxy ($M_* > 5\times 10^6 M_\odot$) and calculate the galaxy abundance around $M_* = 10^6 M_\odot$ with cosmic variance estimated based on volume of $(24 \rm{Mpc}/\it{h})^3$. We speculate that JWST should be able to distinguish FDM models $m_{22} \gtrsim 5$ with CDM by measuring the GSMFs down to masses  $\sim 10^6 M_\odot$ where the difference in galaxy population from different DM models becomes more significant.
}


Although the GSMF is straightforward to predict from hydrodynamical simulations, the galaxy LFs are the direct observable. 
Also, recent measurements of UV LF have extended further into the faint end of the galaxy population using lensed fields, which is important for constraining the FDM model.
Therefore, in the next section, we examine our model predictions in a converse way: 
we derive UV magnitude of galaxies from the simulations and directly compare with the observed LFs.

\begin{figure*}
\includegraphics[width=2.2\columnwidth]{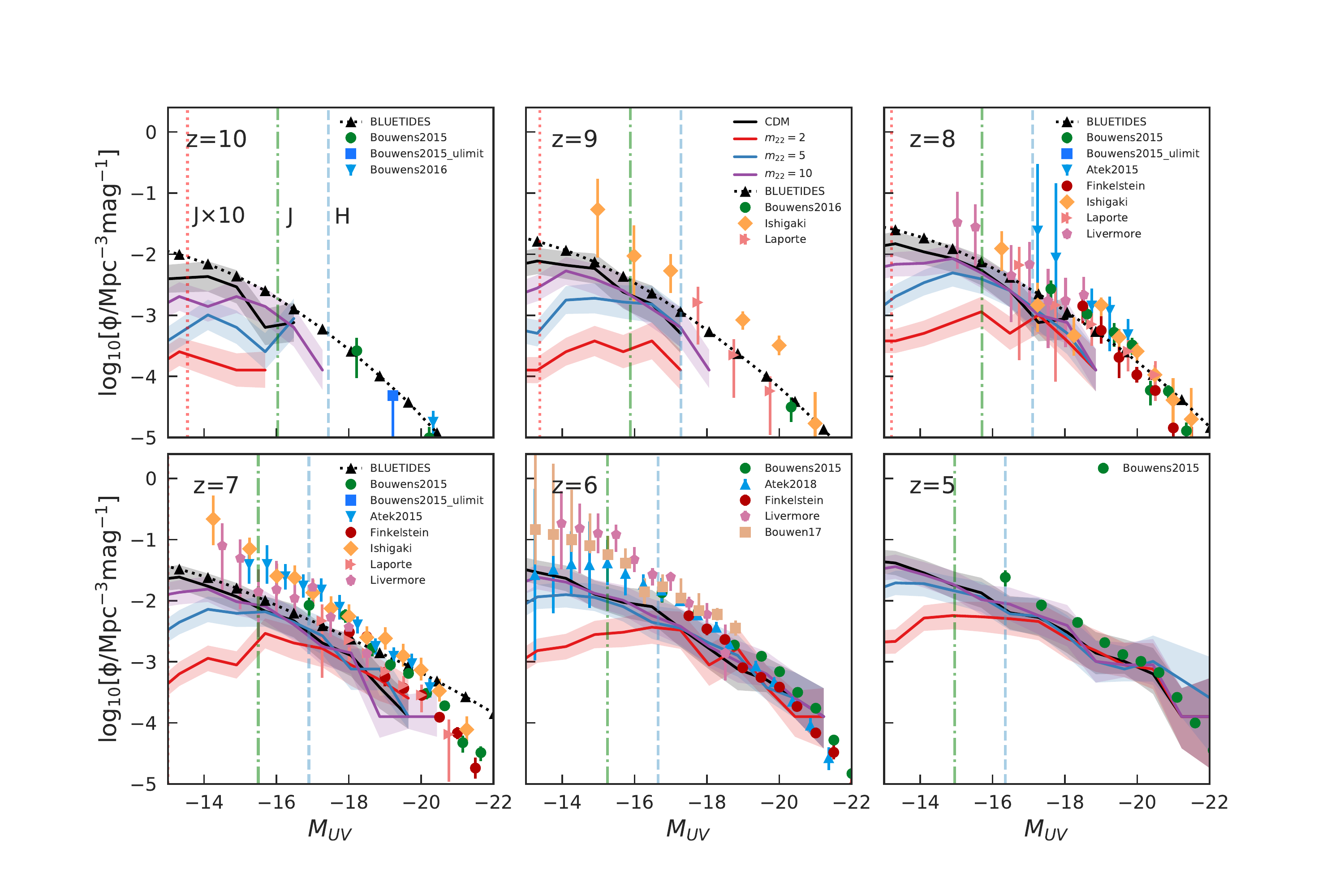}
    \caption{
    LFs for CDM (black solid lines) and FDM models with $m_{22}$= 10 (purple lines), 5 (blue lines), and 2 (red lines) at redshift $z= 5-10$. 
    The colored shaded areas are 1$\sigma$ error cosmic variance for a volume of 15 Mpc/$h$ side length estimated based on \textsc{BlueTides} simulation results.
    The black dotted lines are the results of the \textsc{BlueTides} simulations. 
    The blue vertical dashed line is the detection limit for HUDF.
    Green vertical dash-dotted lines represent the detection limit expected for a JWST deep-field with survey volume comparable to the HUDF.
    Red vertical dotted lines represent the expected $M_{\rm UV}$ limit for JWST lensed fields with 10 $\times$ magnification.
    (The apparent magnitudes of the three lines are $\rm m_{AB,lim}$ = 30.0, 31.5, and 34 respectively).
    For $z=6-9$, the data (at $M_{\rm UV} \geq -17$) is from the analysis of HFF program applying the gravitational lensing techniques \citep{Atek2015,Atek2018,Livermore2017,IshigakI2018}. We use Livermore's Eddington-corrected data points shown in \citet{Yung2019}. }
    \label{fig:uvlf}
\end{figure*}

\begin{figure}    
\includegraphics[width=1\columnwidth]{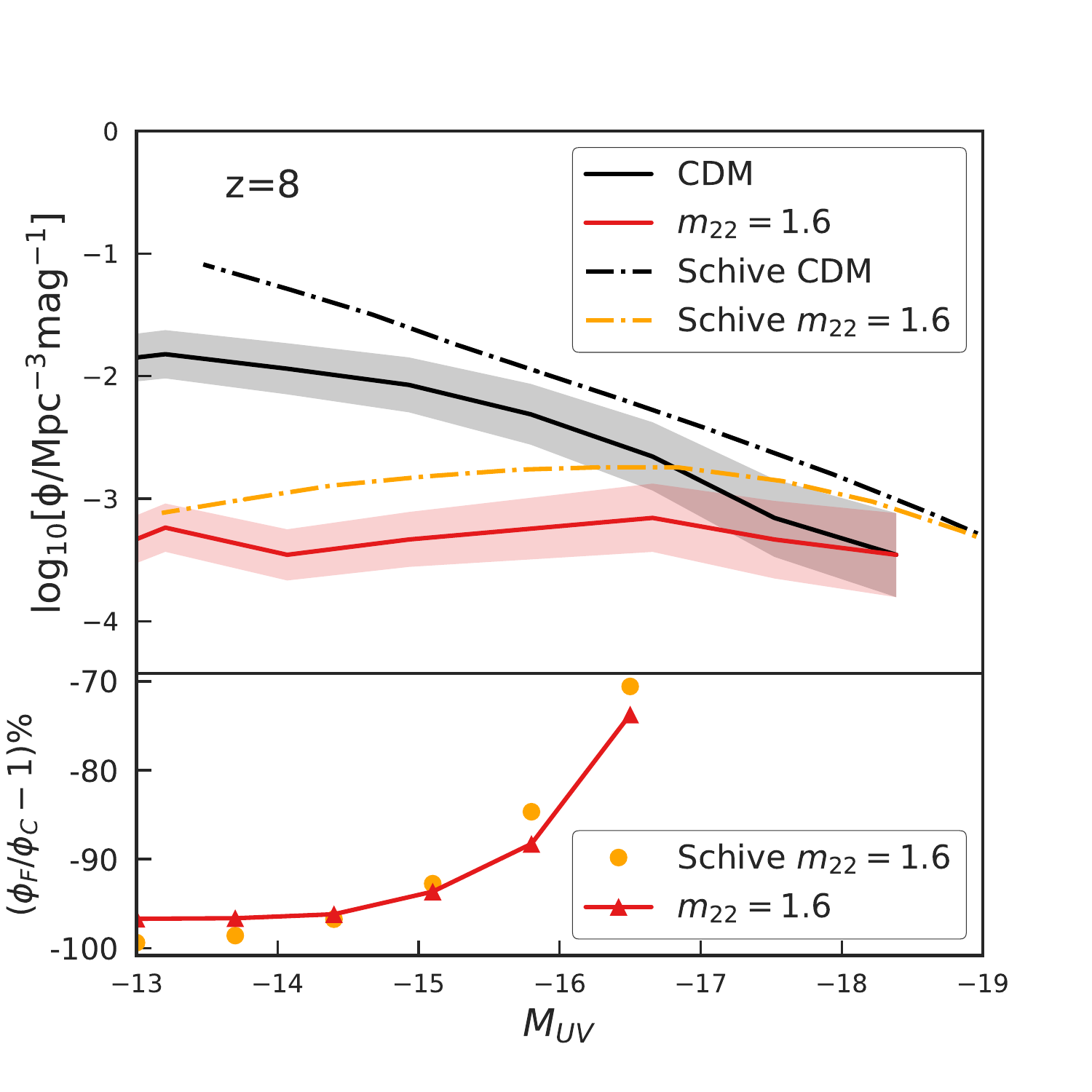}
\caption{The comparison between our predicted LFs and those from \citet{Schive2015} at $z=8$. 
\textit{Upper panel}: 
Black and orange dash-dotted lines are the \citet{Schive2015} results for CDM and FDM with $m_{22}=1.6$ respectively. 
They are obtained by applying conditional LF model on their halo mass function. 
Black and red solid lines are the predictions from our hydrodynamical simulation with the same FDM bosonic mass. 
The color-shaded areas are the estimated $1\sigma$ cosmic variance on volume of $(15 \rm {Mpc/\it{h})^3}$.
\textit{Lower panel}: The ratio of FDM LFs over their CDM counterparts from this work and the \citet{Schive2015} results.}
\label{fig:compare}
\end{figure}


\subsection{Galaxy Luminosity Function}
\label{subsection:UVLF}
During the last decades, rapid progress has been made and increasing numbers of 
high$-z$ faint galaxies have been discovered. Much of this progress has come with the installations of new instruments, such as HST/WFC3 \citep{Windhorst2011}, and the many associated extremely deep surveys that have been carried out, e.g. Cosmic Assembly Near-IR Deep Extragalactic Legacy Survey \citep[CANDELS;][]{Grogin2011}, GOODS \citep[][]{Giavalisco2004}, HUDF \citep[][]{Beckwith2006}, the Cluster Lensing And Supernova survey with Hubble \citep[CLASH;][]{Postman2012}, the Early Release Science field \citep[ERS;][]{Windhorst2011}, and the Brightest of Reionizing Galaxies Survey \citep[BoRG;][]{Trenti2011}. 
Also, applying techniques such as lensing magnification through galaxy clusters have yielded intriguing results for the faint-end LF observations. In particular, the HFF program has been able to identify sources that are intrinsically fainter than the limits of the current unlensed surveys, extending the observed UV LFs to $M_{\rm UV} \sim$ -13 $-$ -15 \citep{Atek2015,Livermore2017,IshigakI2018}.

Fig.~\ref{fig:uvlf} shows the LFs of CDM (black lines) and the three FDM models (with purple, blue, and red solid lines for FDM with $m_{22}$ = 10, 5, and 2 respectively) at $z = 5 - 10$. 
Shaded areas represent the expected 1$\sigma$ cosmic variance uncertainties of comoving volume with 15 Mpc/$h$ side length for each DM model.
Black dotted lines with black triangle points show the results from the \textsc{BlueTides} simulations for comparison.
We can see that there is a marked suppression of faint-end LFs in the FDM models and the suppression becomes more prominent for decreasing values of the FDM boson mass, $m_{22}$.
If we take $z=6$ as an example, the galaxy abundance is suppressed by $> 50\%$ for $M_{\rm UV} \gtrsim -15$ ($M_{\rm UV} \gtrsim -17$) for FDM $m_{22}=5$ ($m_{22}=2$) model.
To trace the time evolution of the suppression, we show (in the lower panel of Fig.~\ref{fig:z_evolution}) the ratio (at the faint-end) of the galaxy number density between CDM and FDM models (in different UV magnitude bin) as a function of redshift. 
Similar to the effects on the GSMF (galaxy number density binned in mass), the suppression of faint-end LFs decreases with decreasing redshift. For example, in the FDM model with $m_{22} =5$ (in blue) model, the abundance of galaxies within magnitude range of $-16 < M_{\rm UV} < -14$ is suppressed by $\lsim 60\%$ at $z=9$ compared to CDM, but the suppression is only $\lsim 20\%$ at $z=5$.

In Fig.~\ref{fig:compare}, we compare the LFs from our simulations to those from \cite{Schive2015}, which were derived by applying a conditional LF model on DM halo mass function constructed from cosmological simulations (DM-only). 
In particular, they employed the least $\chi$-square fitting on the observational data by \cite{Bouwens2015} to determine the best-fit parameters for each conditional LF model applied to the different DM scenarios.
In Fig.~\ref{fig:compare} we show their predicted LFs for CDM and FDM with $m_{22}=1.6$ (with black and orange dash-dotted lines) and compare these with the LFs generated by our direct hydrodynamical simulations (black and red solid lines respectively). 
Our predictions of LFs, either for CDM or FDM, are somewhat systematically lower than the \cite{Schive2015} models. 
Nevertheless, in the lower panel of Fig.~\ref{fig:compare}, where we show the comparison of the ratio of FDM over CDM number density in both studies, we can see that the amount the suppression as a function of magnitude is in good agreement with each other within 5$\%$. 
Therefore both models have captured similar effects of FDM with regards to suppressing the galaxy LFs. 
In essence, the results from full hydrodynamical simulations of galaxy formation predict LFs that do not differ systematically from the \cite{Schive2015} predictions.
This is a promising result, implying that the LF predictions 
from FDM are somewhat stable against different galaxy formation models.


\begin{figure}
	\includegraphics[width=1.1\columnwidth]{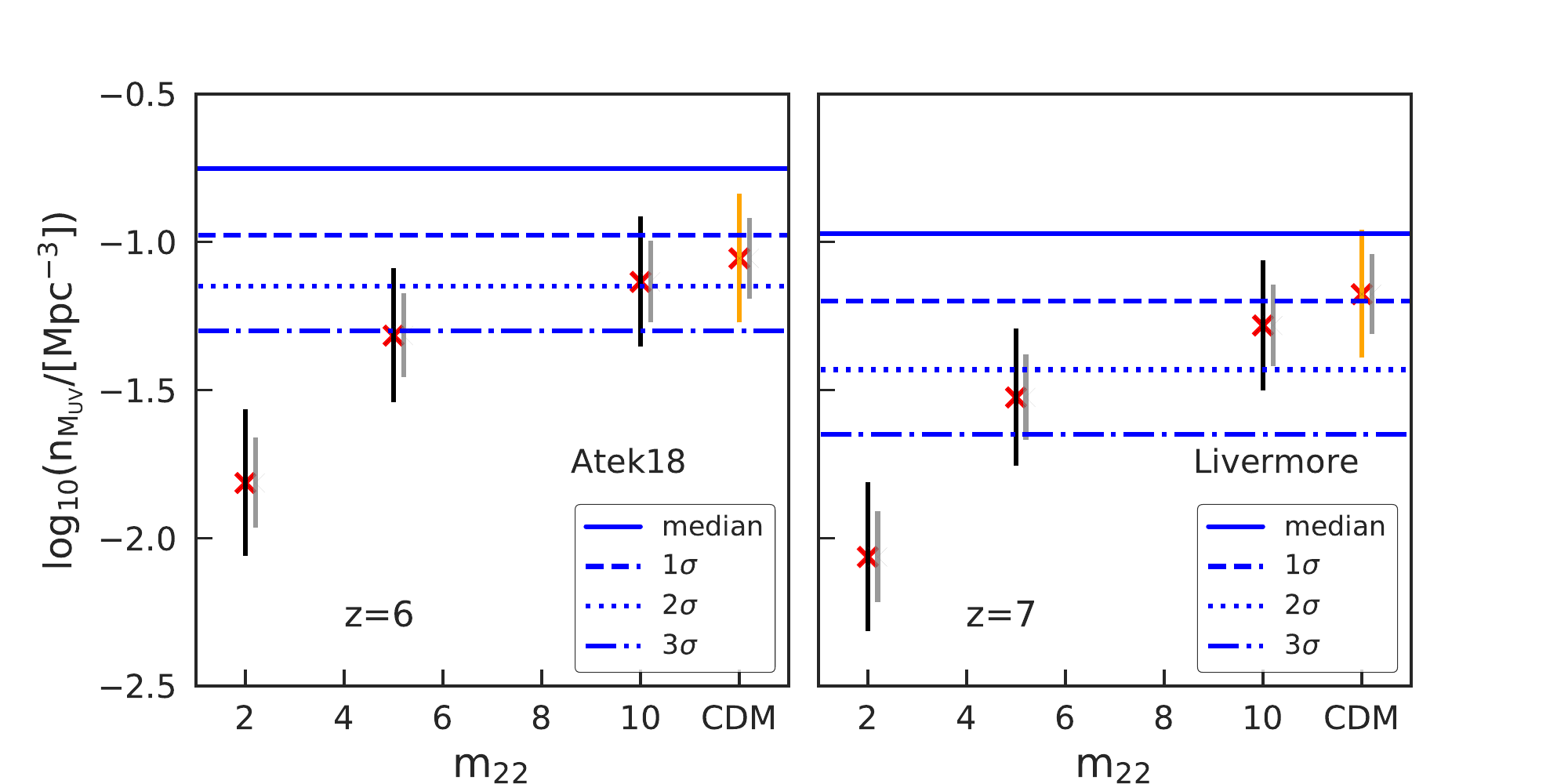}
    \caption{
    Cumulative number densities of galaxies for $\rm -18 \leqslant M_{UV} \leqslant -14$ at $z=6$ (left panel) and at $z=7$ (right panel) as function of $m_{22}$. The error bars correspond to the $1\sigma$ cosmic variance of effective volume of $\rm (15 Mpc/\it{h})^3$ (grey or orange error bars) and $\rm (24 Mpc/\it{h})^3$ (light grey error bars) both calculated from \textsc{BlueTides}.
    The horizontal lines represent the observed cumulative galaxy number density within the same magnitude band with $1\sigma$ (dash lines), $2\sigma$ (dotted lines), and $3\sigma$ (dash-dotted lines) confidence levels. The data set used is from \citet{Atek2018} for $z=6$ and \citet{Livermore2017} with Eddington correction for $z=7$.
    }
    \label{fig:ndensity}
\end{figure}


\subsubsection{Comparison with observations}
The observational data shown in Fig.~\ref{fig:uvlf} have been collected from multiple studies.
For a brief summary here, 
\cite{Bouwens2015} utilizes data sets from CANDELS, HUDF, ERS, and the BoRG/HIPPIES programs, and \cite{Finkelstein2015} uses data sets from CANDELS/GOODS, HUDF, the HFF parallel fields near cluster MACS J0416.1-2403 and Abell 2744.
\cite{Livermore2017} combines HFF data of the Abell 2744 and MACS J0416.1-2403 clusters,
and in this paper we refer to the Eddington corrected version of the \cite{Livermore2017} data shown in \cite{Yung2019}.
\cite{Laporte2016} combines Hubble and ${\it Spitzer}$ data from MACS J0717.5+3745 cluster and its parallel field.
\cite{Bouwen2017} uses galaxy samples from four most massive clusters in HFF to provide prediction for LF at $z \sim 6$ while \cite{Atek2018} combines data sets of all 7 clusters from the HFF program. 
In \cite{IshigakI2018} they utilize the complete HFF data.
For the constraints on the faintest populations ($M_{\rm UV} \geq -17$), where the discrepancies between different DM models become distinctive, all of the observational data sets available now \citep[e.g.,][]{Atek2015,Atek2018,Livermore2017,IshigakI2018} are obtained by taking advantage of lensing magnification by the massive foreground galaxy clusters. 
This technique has the great potential to probe further into the faint-end LFs, however, it also suffers from significant systematics \citep[e.g., see discussions in][]{Menci2017}.
As shown in Fig.~\ref{fig:uvlf}, there are sometimes some discrepancies between different observations at several redshifts (e.g., $M_{\rm UV}\gtrsim -15$ at $z=6$) in the faint-end of the measured LFs (even though all measurements agree with each other within 1 $\sigma$ due to large observational uncertainties).
\cite{Bouwen2017} shows that for highly-magnified sources the systematic uncertainties can imply up to orders of magnitudes differences. 
Therefore the current constraints in range of $M_{\rm UV} \gtrsim -15$ at $z > 4$ are still under significant debate.

Despite the large uncertainties, the LF data from the lensed fields can extend to fainter $M_{\rm UV}$ thresholds than those from the unlensed data, such as those used by \cite{Song2016} to infer the GSMFs. Therefore we expect that some of the LF observations~(especially those which adopted the lensed samples to reach $M_{\rm UV} \gtrsim -17$) can place important constraints on FDM models. 
Using a similar procedure described in the previous section, 
we compare the cumulative galaxy number density over certain magnitude band $\rm n_{obs}$ derived from observations with our DM model predictions (shown in Fig.~\ref{fig:ndensity}) at $z=6$ and $z=7$.
Although the discrepancies between FDM and CDM model predictions increases with higher redshift, the constraint on FDM does not improve significantly by using data sets from $z>7$ due to the large observational uncertainties at higher redshifts (and/or complete lack of data otherwise). So we will restrict our analysis to $z=6, 7$ and calculate $\rm n_{obs}$ using the magnitude range of $\rm -18 \leqslant M_{UV} \leqslant -14$.
At $z = 6$ we do not extend down to $M_{\rm UV}>-14$ again due to the large uncertainties in the observational data sets.

For the purpose of this analysis,
we choose the data sets that are the most consistent with our CDM prediction as those allow us to quantify the constraints on FDM models more reliably.
In Fig.~\ref{fig:ndensity}, the horizontal lines indicate the median (the solid lines), $1\sigma$, $2\sigma$, and $3\sigma$ lower bounds (the dash, the dotted, and the dash-dotted lines) of $\rm n_{obs}$ obtained from \cite{Atek2018} at $z=6$ (left panel) and \cite{Livermore2017} (Eddington-corrected version from \cite{Yung2019}) at $z=7$ (right panel). 
The data points marked by the red cross are cumulative number densities ($n_{\rm M_{UV}}$) obtained from different FDM simulations, while CDM models are indicated by the rightmost points.
Comparing the predicted $n_{\rm M_{UV}}$ and the corresponding lower bounds derived from \cite{Atek2018}, we can see that $m_{22}=5$ model is ruled out at the $3\sigma$ confidence level at $z=6$.


We evaluate the expected cosmic variance on $n_{\rm M_{UV}}$ using the \textsc{BlueTides} simulation based on simulation subvolumes with size of $\rm {(15 Mpc/\it{h})}^3$ and $\rm {(24 Mpc/\it{h})}^3$, which is comparable to the effective volume of the HFF lensing fields \citep{Livermore2017} and the HUDF respectively. 
They are marked by error bars (1$\sigma$) in Fig.~\ref{fig:ndensity}. For example at $z=6$, the corresponding $1\sigma$ cosmic variance is $\rm \sigma_{log(n_{M_{UV}})} = 0.22$ and 0.14 for these two different volumes for the CDM model.
It is worth noticing that, although the cumulative number densities from our CDM realizations is slightly lower than the observed median values, they agree well with observations within $1\sigma$ (for both with effective volume comparable to HFF or HUDF) when we take into consideration the cosmic variance. This indicates that our CDM models provide a good baseline for quantifying effects of FDM on the observed galaxy number densities. 
However, when checking LFs alone, our CDM LFs are generally lower than observations, especially for $M_{\rm UV} \geq -16$ where observational data are obtained from highly magnified regions.

We note that in Fig.~\ref{fig:ndensity} the $n_{\rm M_{UV}}$ for CDM and FDM $m_{22}=10$ models are almost indistinguishable within the $1\sigma$ cosmic variance for both effective volumes. Observational data with improved precision, and extending to fainter magnitudes, is needed before further constraints on the FDM bosonic mass can be achieved. Larger survey effective volumes would also be required to robustly distinguish between different models.

In Fig.~\ref{fig:uvlf}, the vertical blue dashed lines show the detection limits of HUDF \citep[with apparent magnitude $\rm m_{AB,lim} =30.0$,][]{Koekemoer2013}, 
the green vertical dash-dotted lines represent the limits expected for JWST NIRCam filter with HUDF-like deep field survey ($\rm m_{AB,lim} = 31.5$), 
and red vertical dotted lines represent the expected $M_{\rm UV}$ limit for JWST lensed fields with $10\times$ magnification ($\rm m_{AB,lim} = 34$) \citep{Yung2019}.
JWST's capabilities should allow more secure measurements of photometric redshifts for high$-z$ candidates and probe further the faint end of LFs. 
It will therefore provide more solid observations for the faint galaxy number densities and therefore allow for tighter constraints on FDM models.
To roughly estimate JWST's potential for this problem, we estimate how well CDM and FDM scenarios can be distinguished from each other at the magnitude limit expected for a JWST-like survey. 
We take the expected limit $M_{\rm UV} \sim -13$ for JWST lensed fields with $10\times$ magnification, and calculate galaxy abundance around this limit assuming that uncertainties are dominated by cosmic variance with the effective volume of $\rm (24 Mpc/\it{h})^3$ (comparable to HUDF). We find that the UV LF observations at $z \geq 7$ have the power to distinguish FDM $m_{22}=10$ model from the CDM scenario by 2$\sigma$, whereas at lower redshift ($z \leqslant 6$) the model discrepancies decrease so that larger survey volumes are needed to reduce uncertainties.

\begin{figure}
    \includegraphics[width=\columnwidth]{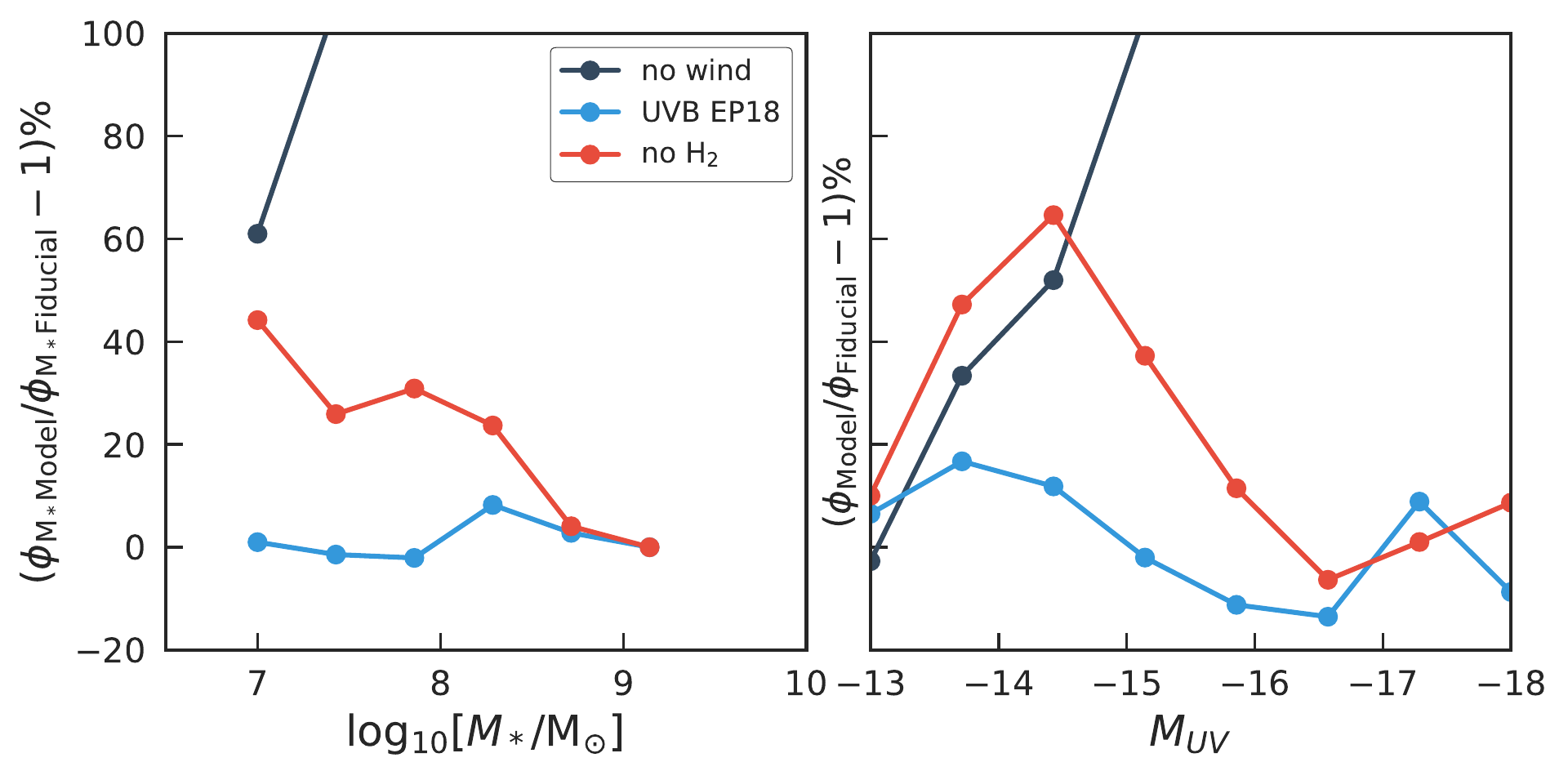}
    \caption{The effect of different sub-grid models on GSMFs (left panel) and UV LFs (right panel) at $z=6$.
    Black lines show the results of turning off the stellar wind feedback. 
    Blue lines represent the results of simulations adopted the UV background model from \citet{Puchwein2018} .
    Red lines indicate the results with $\rm H_2$-based star formation model turned off.
    }
    \label{fig:subgrid}
\end{figure}

\subsubsection{Possible effects from the sub-grid modeling}
\label{subsubsetion:sub-grid}
The FDM model implementations we have adopted here are fully consistent with those
adopted in previous works which consider high-z galaxy abundance calibrated by DM-only simulations \citep{Schive2015, Menci2017}. Having introduced the hydrodynamics and galaxy formation models, we now have a fully self-consistent set of simulations that predict the LF directly, under reasonable assumptions about baryonic physics (for the given resolution regime).
This provides interesting constraints on the FDM bosonic mass. 


\yueying{
However, we also note that it is possible that the predictions for the UV LFs might be affected by different sub-grid physical models.
For example, recently \cite{Yung2019} have explored how different galaxy formation recipes (in semi-analytic models) can influence faint-end LFs.
In this section, we broadly test some of the effects of a few different sub-grid models and examine whether the details of sub-grid models may potentially lead to weaker constraints on FDM.
}

Several simulation studies \citep[e.g.][]{Jaacks2013,Thompson2013,Oshea2015} have shown that the predicted faint-end LFs are actually lower for the $\rm H_2$-based star formation models, such as the one we have adopted, than those from models applying a uniform threshold density for star formation.
To assess to what extent this may affect our predictions, we run a test simulation where we completely remove the $\rm H_2$ formation.
\yueying{In Fig.~\ref{fig:subgrid}, we show the fractional differences of GSMFs (left panel) and UV LFs (right panel) between the fiducial model and the run without $\rm H_2$-based star formation in red lines.}
Indeed, we find that removing $\rm H_2$-based star formation model leads to about 40-20\% enhancement to the faint galaxy populations, while for brighter galaxies, e.g. $M_* \gsim 5 \times 10^8 M_\odot$, the predictions remain the same. These results agree with what was seen in previous tests using cosmological SPH simulation code \citep{Thompson2013}.
Obviously, removing $\rm H_2$-based star formation does not provide a better physical model for the galaxy population.
Several recent high-resolution simulations of single, zoomed halos have started to adopt more realistic models for molecular-based star formation. However, what they typically find is that molecular-based star formation produces an enhancement in star formation rates for galaxies at stellar masses below $10^{7} {\rm M_\odot}$ \citep[see Fig. 7 in][]{Jaacks2018}. 
This is a regime we are not able to simulate with our current large uniform volume simulations. But with upcoming JWST pushing the current detection limits to fainter magnitudes, it will become increasingly important to carry out simulations with more realistic molecular-based star formation models.

Another possible element of the sub-grid physics that can lead to a suppression in the LFs is the stellar wind feedback. 
Again, for the sake of examining the scales at which stellar winds influence in the sub-grid models, we turn off this feature in another test run. 
\yueying{Black lines in Fig.~\ref{fig:subgrid} give the fractional differences between the fiducial run and the test run that turns off the stellar feedback.} 
We find that this brings huge enhancement to galaxy abundances which has already been ruled out by current observations at the bright-end. We note that stellar feedbacks are typically more efficient at brighter-end, which can be seen from Fig.~\ref{fig:subgrid}.
Naturally, more realistic models for stellar winds can affect the low-mass end of GSMF, and correspondingly, the faint-end of the LFs.
However, note that the level of influence of stellar wind feedback is strongly constrained by bright-end galaxy LF constraints. Therefore, it appears to have a relatively small effect on the stellar mass and luminosity region where FDM significantly suppresses the galaxy population.

Finally, we examine whether our choice of UV background affects the faint galaxy population. We apply another commonly used UV background synthesis model from \cite{Puchwein2018}, which adopts a much lower (by about three orders of magnitude) photo-ionization and photon heating rate at $z > 6$ \citep{fg09} for comparison.
\yueying{We give the fractional difference due to UV background changes in blue lines in Fig.~\ref{fig:subgrid}.}
We find that the specific UV background model used here has no prominent effects ($< 20\%$) on galaxy formation at the scales of interest.

We conclude that the uncertainties lying in some aspects of the sub-grid models may play a role in loosening the constraints on FDM models
(as some could plausibly enhance the faint-end UV LF). 
However, the effects are likely not sufficient (when reasonable models are taken into account that provide a good fit both to the faint and the bright-end of LFs) to compensate the suppression brought about by FDM models around our current limits, which is $m_{22} \lsim 5$.
For example, if we assume that our simulated FDM LFs are enhanced by $\sim 50\%$ due to different star-formation models, our FDM limit of $m_{22} > 5$ obtained from Fig.~\ref{fig:ndensity} will only be slightly degraded to $m_{22} > 4$.

\section{Conclusions}
\label{section4: Conclusion}

In this work we generate, for the first time, full hydrodynamical cosmological simulations of galaxy formation within the framework of FDM models to examine the predictions of the high-redshift galaxy abundance. 
Our fiducial galaxy formation model is built upon the successes of the \textsc{BlueTides} simulation that has been validated against high-redshift galaxy population observations \citep{feng2016,Wilkins2017,Wilkins2018, Bhowmick2018}.
We have run a suite of FDM simulations with FDM bosonic mass ranging from $m_{22}=1.2 \sim 10$.
We have shown that small-scale structure is washed out in FDM scenario, and the resulting suppression on galaxy abundance becomes more prominent for lower values of $m_{22}$.
Using the current and future high-redshift UV LF observations, we are able to constrain a lower limit for $m_{22}$.
Our main results can be summarized as below:

\begin{itemize}
\item In terms of halo MFs, the FDM halo abundance is suppressed by $\gtrsim 50\%$ in the halo mass range $M_h \lsim 5\times 10^{10} M_\odot$ for $m_{22}=5$ ($\lsim 10^{10} M_\odot$ for $m_{22}=2$).

\item Regarding how galaxies populate the DM haloes in FDM scenarios, 
we have found that due to the delayed star formation process in FDM models, (which is rather prominent for small mass haloes ($\rm M_h \lsim 10^{10} M_\odot$),
FDM galaxies residing in those halos have a lower luminous fraction and higher UV band luminosity comparing to their CDM counterparts. 
For example, we found that in FDM $m_{22}=2$ model at $z=6$, the fraction of halos with $\rm M_h \lsim 10^{10} M_\odot$ that host galaxy is $80\%$ smaller than that in CDM. However, the age of galaxies is about $60\%$ younger than that in CDM and the corresponding rest-frame UV luminosity is about $0.5 \sim 1$ magnitude brighter. 

\item Our predictions for GSMFs showed that the discrepancies between CDM and FDM models become significant for $\rm M_* \lesssim 10^7 M_{\odot} (\lesssim 10^8 M_{\odot})$ for FDM models with $m_{22} = 5$ ($m_{22} = 2$); 
the discrepancies decrease with decreasing redshift.
The abundance of galaxies ($\rm M_* \sim 10^7 M_{\odot}$) is suppressed by $\sim 40\% (80\%)$ at $z = 9$, $\sim 20\%$ ($60\%$) at $z = 5$ for FDM model with bosonic mass $m_{22} = 5$ ($m_{22} = 2$). 
For galaxies with stellar masses in the range $\rm 10^7 M_\odot < M_* < 10^9 M_\odot$, a survey volume comparable to HUDF is not capable to distinguish between $m_{22}=5$ and CDM model using GSMFs when the cosmic variance is considered.
Larger survey volumes but still able to extending to faint/small mass end will be necessary to move the constraints to higher values of $m_{22}$ in FDM.

\item Recent observations of UV LF based on the lensed fields of probe faint-end LF that is beyond the detection limit of the unlensed HUDF, and thus place tightest constraints on FDM. 
By examining galaxy populations in UV magnitude band, we find that suppression in LFs due to FDM models becomes prominent at the magnitude range of $M_{\rm UV} \gtrsim -17$, and the difference reduces with decreasing redshifts. 
In particular, the number density of faint-end galaxies in the magnitude band of -16 < $M_{\rm UV}$ < -14 is suppressed by $\lsim 60\%$ ($90\%$) at $z=9$, $\lsim 20\%$ ($70\%$) at $z = 5$ for FDM model with bosonic mass $m_{22} $ = 5 ($m_{22}$ = 2). 
Comparing with the cumulative number densities derived from \cite{Atek2018} at z = 6 in magnitude range of $\rm -18 \leqslant M_{UV} \leqslant -14$, we are able to rule out $m_{22}$ = 5 model with $3\sigma$ confidence level. 

\end{itemize}

Comparing our results with previous work, we find that our constraints are overall consistent with previous FDM predictions for faint-end LFs. 
For example, \cite{Schive2015} uses DM-only simulations to model the structure formation in FDM universe and applied conditional LF model to infer galaxy LFs. 
Using this approach, they gets a lower bound of bosonic mass $m_{22} > 1.2$ ($2\sigma$ confidence level). 
Using the FDM halo mass functions and empirical model for the LFs developed by \cite{Schive2015}, \cite{Menci2017} has recently extended the lower bound constraint to $m_{22} > 8$ (by $3\sigma$) by comparing FDM halo abundance with cumulative galaxy number density inferred from LF observations by \cite{Livermore2017}.
This, however, was the original faint-end LF measurement from \cite{Livermore2017} (with no Eddington correction) which is very high and hence leads to a stronger constraint on FDM.
Also, \cite{Menci2017} limits are perhaps optimistic as they rely on an extrapolation of the analytic fitting function on FDM halo MF which was originally fitted in the range of $0.8 \leq m_{22} \leq 3.2$ \citep{Schive2015}.


Considering the fact that the prediction for the faint-end of the galaxy LFs might be sensitive to different sub-grid models adopted in the hydrodynamic simulations,
we have broadly examined some possible effects of subgrid physics modelling on our predictions.
For example we have tried disabling our $\rm H_2$-based star formation model, turning off stellar wind feedback, and changing UV background models to check the effects on the faint-end UV LF.
Although none of these represent an improvement in the physical modeling, we use them to gain insights on the potential sensitivities of FDM limits to these effects.
We find that although some of them exhibit effects $\sim$ 20$\%$ or higher on the GSMF or UV LF, apart from the $\rm H_2$ models~(which only affects the faint-end of the galaxy population), others can be tightly constrained by observations of galaxy abundances at the bright-end. Future work will be needed to assess fully the effects of sub-grid physics in FDM models.

Similar to WDM models, FDM simulations suffer from the spurious fragmentation which is caused by the cutoff in primordial power spectrum and brings artificial enhancement to small halo abundance.
\yueying{However, we have demonstrated that FDM halos that host luminous galaxies are more massive than the halo mass region where spurious halos resides.}
Therefore, for hydrodynamical simulations, we do not need to rely on an accurate correction for spurious haloes since we only focus on the impact of FDM models on galaxy properties.

Our FDM simulation predicts lower luminous fraction for small mass haloes ($\rm M_h \lsim 10^{10} M_\odot$).
It is worth noticing that different DM models lead to different behaviour in terms of halo luminous fraction.
For example, in \cite{Lovell2017} where they perform hydrodynamical simulations in the ETHOS DM scenario (which also has similar cut-off feature in primordial power spectrum but with additional DM self-interaction), they find a reversed trend in the luminous fraction whereby small mass halos in ETHOS have a higher luminous fraction than in CDM.
They speculate that this is due to the interplay of monolithic collapse in ETHOS structure formation together with the heating of stellar feedback.
However, the luminosity for the small mass halos has similar trend as our FDM result: in ETHOS galaxies $M_{\rm U} \sim 0.3$ mag brighter for small mass haloes ($\rm M_h \lsim 10^{10} M_\odot$).

In this study, we do not implement QP of FDM in our non-linear evolution.
Recent work has been carried out  to take into account the effect of QP term on structure formation \citep[e.g.,][]{Zhang2017} and shown that the QP brings further suppression of $\sim 10\%$ on halo abundance around $M_h \sim 10^{10} M_\odot$ at $z=5$, and $\sim 7\%$ at $z=10$ (Jiajun Zhang, private communication). 
These effects are much smaller than any of the uncertainties in the current faint-end UV LF observations or those coming from the sub-grid  models. 
We note that by ignoring QP in our method we would slightly overestimate the halo abundance at low-mass end, which would end up giving more conservative constraints on FDM bosonic mass.

Currently there are still large uncertainties lying in the UV LF observations within the lensed fields which are the ones that probe the faintest galaxies at high redshift.
Upcoming observations with JWST will be crucial for improving measurements at the faint-end LFs and put further constraints on FDM models. 
From our UV LF predictions, we expect that at depths of $M_{\rm UV} \sim -13$~(close to detection limit of JWST), and with cosmic variance from a survey volume comparable to the HUDF, we will be able to distinguish FDM model with $m_{22} > 10$ from CDM scenario at 2$\sigma$ confidence level at redshift $z \geq 7$.

\section*{Acknowledgements}
We thank Jiajun Zhang for useful discussions.
We thank Aklant Bhowmick and Kuan-Wei Huang for the help with this project.
We acknowledge funding from NSF ACI-1614853, NSF AST-1517593, NSF AST-1616168, NASA ATP 80NSSC18K1015
and NASA ATP 17-0123.
MYW acknowledges support of the McWilliams Postdoctoral Fellowship.
The \texttt{BLUETIDES} simulation was run on the BlueWaters facility at the National Center for Supercomputing Applications

\bibliographystyle{mnras}
\bibliography{bib.bib}


\bsp	
\label{lastpage}

\end{document}